\documentclass[review]{elsarticle}

\usepackage{lineno,hyperref}
\usepackage{array}
\usepackage{amssymb}
\usepackage{enumerate}
\usepackage{makecell}
\usepackage{mathtools}
\usepackage{array}
\usepackage{multirow}
\usepackage{color}
\usepackage{diagbox}
\usepackage{enumerate}
\usepackage{graphicx}
\usepackage{subfigure}
\usepackage{CJK}
 \usepackage{tikz,mathpazo} 
\usepackage{indentfirst}
\usepackage{soul}
\usepackage{amsmath}
\usepackage[justification=centering]{caption}
\usepackage{tabularx}
\usepackage{algorithm}
\usepackage{algorithmicx}
\usepackage{algpseudocode}
\usepackage{adjustbox}
\usepackage{geometry}
\usetikzlibrary{shapes.geometric, arrows}
\modulolinenumbers[5]
\newtheorem{definition}{Definition}[section]
\newtheorem{example}{Example}[section]
\newtheorem{proposition}{Proposition}[section]

\soulregister{\cite}7
\soulregister{\ref}7

\newcommand{\tabincell}[2]{\begin{tabular}{@{}#1@{}}#2\end{tabular}}
\newcommand{\circledsmall}[1]{\lower.7ex\hbox{\tikz\draw (0pt, 0pt)%
    circle (.5em) node {\makebox[0.1em][c]{\small#1}};}}

\newcommand{\circledtiny}[1]{\lower.7ex\hbox{\tikz\draw (0pt, 0pt)%
    circle (.3em) node {\makebox[0.1em][c]{\tiny #1}};}}

\journal{Journal of \LaTeX\ Templates}

\begin{document}

\begin{frontmatter}

\title{Fractal-based Belief Entropy}

\author[a]{Qianli Zhou}
\ead{zhouqianli@std.uestc.edu.cn}

\author[a,b,c,d]{Yong Deng\corref{mycorrespondingauthor}}
\cortext[mycorrespondingauthor]{Corresponding author}
\ead{dengentropy@uestc.edu.cn}
\address[a]{Institute of Fundamental and Frontier Science, University of Electronic Science and Technology of China, Chengdu 610054, China}
\address[b]{School of Eduction Shaanxi Normal University, Xi’an, 710062, China}
\address[c]{School of Knowledge Science, Japan Advanced Institute of Science and Technology, Nomi, Ishikawa 923-1211, Japan}
\address[d]{Department of Management, Technology, and Economics, ETH Zurich, Zurich, 8093, Switzerland}

\begin{abstract}
The total uncertainty measurement of basic probability assignment (BPA) in Dempster-Shafer evidence theory (DSET) has always been an open issue. Although some scholars put forward various measurements and entropies of BPA, due to the existence of discord and non-specificity, there is no method can measure BPA reasonably. In order to utilize BPA to  practical decision-making, pignistic probability transformation of BPA is a significant method. In the paper, we simulate the pignistic probability transformation (PPT) process based on the fractal idea, which describes PPT process in detail and shows the process of information volume changes during transformation intuitively. Based on transformation process, we propose a new belief entropy called fractal-based belief (FB) entropy. After verification, FB entropy is superior to all existing belief entropies in terms of total uncertainty measurement and physical model consistency.
\end{abstract}

\begin{keyword}
Dempster-Shafer evidence theory \sep Fractal \sep Belief entropy \sep Pignistic probability transformation \sep Total uncertainty measurement 
\end{keyword}

\end{frontmatter}

\linenumbers
\section{Introduction}
\label{intro}

Dempster-Shafer evidence theory (DSET) \cite{dempster2008upper,shafer1976mathematical} as a generalization of probability theory (PT) express the information by interval probabilities. For an $n$-element mutually exclusive set, probability distribution express its information by $n$ probabilities, and in DSET, $2^n$ mass functions called focal elements formed basic probability assignment (BPA) to express information. BPA utilizes more dimensional data than probability distributions, so it has the ability to express more information than probability distributions. Relying on the above advantages, DSET is widely applied in information fusion \cite{Xiong2021InformationSciences,yang2013discounted,pan2020association}, data de-combination \cite{fan2021combination}, reasoning \cite{liao2020deng}, and reliability evaluation \cite{gao2021NET}. Because the power set means the combination numbers \cite{Song2021powerset}, the permutation numbers subset not be considered. Smarandache and Dezert \cite{smarandache2006advances} extended $2^n$ to $U^n$ to propose the Dezert-Smarandache theory (DSmT), which can express more generalized information than DSET, and Xiao  \cite{Xiao2020CEQD,xiao2021caftr} extended BPA to the complex number field to predict interference effects in a more proper way.

In PT, Shannon entropy \cite{shannon2001mathematical} can express the uncertainty of probability distribution, but how to measure the total uncertainty of BPA is still an open issue. BPA can be seen as formed by two properties discord and non-specificity \cite{jousselme2006measuring}. Discord represents the conflict between elements in the framework, and non-specificity, as a difference between BPA and probability distribution, represents the uncertainty of the distribution. In order to facilitate understanding, we divide common BPA measurement methods into $3$ types.

\begin{description}
\item[\textbf{Local measurement: measure a certain characteristic of BPAs}] Hohle \cite{hohle1982entropy} and Yager \cite{yager1983entropy} respectively utilized belief function and plausibility function to calculate the confusion and  dissonance of BPAs. Hartley entropy was proposed to express the non-specificity of BPA in \cite{higashi1982measures}. Klir \textit{et al.} measure the discord of BPA in \cite{klir1990uncertainty}. Harmanec \textit{et al.} \cite{AU} proposed a method to measure the aggregate uncertainty (AU) of BPA. Jousselme \textit{et al.} \cite{jousselme2006measuring} substituted pignistic probability transformation in to Shannon entropy to propose ambiguity measure (AM). We proposed the belief eXtropy to measure the negation degree \cite{zhou2021eXtropy}. 
\item[\textbf{Splitting method: measure uncertainty after splitting the mass functions}] Pal \cite{pal1993uncertainty} first utilized the splitting to divide the mass functions of the focal elements with $n$ elements into $n$ parts and then substituting them into Shannon entropy. Based on above, Deng \cite{Deng2020ScienceChina,Deng2020InformationVolume} splitting the mass functions to their power set, which can represent more uncertainty, and Abell{\'a}n \textit{et al.} evaluated Deng entropy and its extensions in \cite{abellan2017analyzing,moral2020critique}. These two methods satisfy non-negativity, monotonicity, probability consistency, and additivity. However, their maximum entropy distribution is not a vacuous BPA, which is counter intuitive.
\item[\textbf{Belief functions: measure uncertainty based on belief functions}]Due to the limitations of BPA to express information, many measurements use belief functions to express its uncertainty. Wang and Song \cite{wang2018uncertainty} utilized elements' belief (Bel) function and plausibility (Pl) function to measure the discord and non-specificity respectively (Hereinafter referred to as SU measurement). Jirou{\v{s}}ek and Shenoy combined Pl function and Hartley entropy to proposed a new entropy (Hereinafter referred to as JS entropy) \cite{jirouvsek2018new}  and they also proposed a decomposable entropy based on commonality (q) function \cite{jirouvsek2020properties}. Yang and Han \cite{yang2016new} proposed a novel uncertainty measure based on the distance of elements' Bel functions and Pl functions.
\end{description}

There are total $10$ requirements for total uncertainty measurement (UM) methods of BPA in \cite{klir2013uncertainty,abellan2008requirements}. Although some of them are controversial, they can evaluate UM methods comprehensively.

The elements in framework are mutually exclusive, so in process of decision-making, how to transform the BPA to probability distribution is significant. Pignisitic probability transformation (PPT) is utilized in decision layer of transfer belief model (TBM) \cite{smets2005decision}, which distributes the mass functions of multi-element focal elements equally under the principle of keeping the maximum uncertainty. Cobb and Shenoy \cite{cobb2006plausibility} proposed plausibility transformation method (PTM) based on the elements' Pl functions, which has Dempster combination rule consistency. In addition, there are many methods of probability transformation methods \cite{CHEN2021104438} and Han \textit{et al.} evaluated them in \cite{han2015evaluation}. Probability transformation can also be regarded as the non-specificity loss. The previous methods only gave the results of the transformation, and did not describe the process of generating the probability. Therefore, their reasonability only can be evaluated from results , which is not comprehensive enough.

In the paper, we propose a possible PPT generation process based on fractal, and based on this process, we propose a new belief entropy called Fractal-based belief (FB) entropy to measure the total uncertainty of BPA. After evaluation and comparison, we prove that FB entropy can meet the requirements in numerical calculation and has a corresponding physical model. The contributions of paper is summarized as follows: (1) We first use fractal idea to simulate the process of probability transformation. (2) FB entropy can not only measure the uncertainty of BPA reasonably but has corresponding physical model as well, which is superior to all existing belief entropy. (3) We does not deliberately consider discord and non-specificity when defining FB entropy, but we can separate the two parts of uncertainty based on the fractal result. For different BPAs, the proportions of the two parts are different, which is more intuitive. In general, the structure of this paper is as follows:
\begin{description}
\item[$\bullet$]The Section \ref{preliminaries} mainly introduces the basic concepts of DSET, common probability transformation methods and classical uncertainty measurements of BPA.
\item[$\bullet$]In the Section \ref{process}, we simulate the process of PPT based on the fractal and give it a possible explanation.
\item[$\bullet$]Section \ref{fbentropy1} is the core of the paper. According to the process of PPT, we propose the FB entropy. After evaluation its properties, we prove FB entropy can measure BPA rationally.
\item[$\bullet$]Some unique advantages of FB entropy are shown in Section \ref{fbentropy2} by comparing with common methods. 
\end{description}

\section{Preliminaries}
\label{preliminaries}

\subsection{Dempster-Shafer evidence theory}

\begin{definition}[BPA]\label{bpa}\cite{dempster2008upper}
For a finite set $\Theta$ with $n$ elements, it can be written as $\Theta=\{\theta_{1},\dots,\theta_{n}\}$, which is called a discernment framework in DSET. The mass functions of elements in $2^\Theta$ can be written as $\mathbb{B}(2^{\Theta})=\{m(\varnothing),m(\theta_{1}),\dots, m(\theta_{n}), m(\theta_{1}\theta_{2}),\dots, m(\theta_{1}\dots\theta_{n})\}$, and $m(F_i)$ satisfies
\begin{equation}
m(\varnothing)=0;~~~~~~~\sum_{F_{i}\in 2^\Theta} m(F_{i})=1;~~~~~~m(F_{i})\geqslant 0,
\end{equation}
where $\mathbb{B}(2^{\Theta})$ is basic probability assignment (BPA), and $\{F_i\}$ is called focal element.
\end{definition}

This paper only discusses normalized BPA, so $m(\varnothing)=0$. In addition to mass functions, belief functions also can store the information of BPA.

\begin{definition}[Belief functions]\label{beliefinterval}\cite{shafer1976mathematical}
For an $n$-element discernment framework $\Theta$, with its BPA $\mathbb{B}(2^{\Theta})$, the belief (Bel) function, plausibility (Pl) function, and commonality (q) function of focal elements are defined as
\begin{equation}
\begin{aligned}
&Bel(F_{i})=\sum_{G_{i}\subseteq F_{i}}m(G_{i})=1-Pl(\overline{F_{i}}),\\
&Pl(F_{i})=\sum_{G_{i}\cap F_{i} \neq \varnothing~and~G_{i}\subseteq X}m(G_{i})=1-Bel(\overline{F_{i}}),\\
&q(F_{i})=\sum_{F_i\subseteq G_i}m(G_i).
\end{aligned}
\end{equation}
It is obvious that the $Bel(A)\leqslant Pl(A)$, and the belief interval of focal element $A$ is $[Bel(A),~Pl(A)]$.
\end{definition}

The above two methods in Definition\ref{bpa} and \ref{beliefinterval} are usually used to calculate the uncertainty of information in DSET. Next,  some common probability transformation methods are shown.

\subsection{Common probability transformation methods}

\begin{definition}[PPT] \cite{smets2005decision}
\label{PPT}
For an $n$-element discernment framework $\Theta$ with its BPA $\mathbb{B}(2^{\Theta})$. Its pignistic probability transformation (PPT) called $BetP(\theta_i)$ is defined as :
\begin{equation}\label{PPTe}
BetP(\theta_i)=\sum_{\theta_{i} \in F_{i}~and~F_{i}\in 2^\Theta} \frac{m(F_i)}{|F_{i}|} ,
\end{equation}
where $|F_{i}|$ is the cardinality of focal element $F_{i}$.
\end{definition}

\begin{definition}[PTM]\cite{cobb2006plausibility}
\label{PPF}
For an $n$-element discernment framework $\Theta$ with its BPA $\mathbb{B}(2^{\Theta})$. Its plausibility transformation method (PTM) called $PnPl(\theta_i)$ is defined as :
\begin{equation}
PnPl(\theta_i)=\frac{Pl(\theta_{i})}{\sum_{j=1}^{n}Pl(\theta_{j})}
\end{equation}
\end{definition}

Besides PTM, other probability transformation methods are specializations of PPT, i.e., the support of $m(\theta_i)$ plus the support degree of multi-element focal elements for $\theta_i$. Though PTM doesn't satisfy the upper and low probability rule, it is the only method satisfies the Dempster combination rule consistency \cite{han2015evaluation}.

\subsection {Classical uncertainty measurements (UM) of BPA}

\begin{definition}[UM]\label{um}
For a discernment framework $\Theta=\{\theta_{1},\theta_{2},\dots,\theta_{n}\}$, its BPA is $\mathbb{B}(2^{\Theta})$, PPT is $ P_{\mathbb{B}}(\theta_{i})$, and PPF is $Pl\_ P_{m}(\theta_{i})$. Common uncertainty measurements of BPA and its maximum distribution  are shown in Table\ref{d1t1}.
\newgeometry{left=2cm, right=2cm, top=2cm, bottom=2cm}
\begin{table*}[htbp!]\footnotesize
\centering
\begin{adjustbox}{angle=90}
\begin{tabular}{ccccc}
  \Xhline{1.4pt}
Methods & Expression   &\tabincell{c}{Maximum distribution}&Maximum& Remark  \\
  \Xhline{1.4pt}
  \tabincell{l}{Ambiguity measure\cite{jousselme2006measuring}}&\tabincell{l}{$H_{j}=-\sum_{i=1}^{n}P_{\mathbb{B}}^{\Theta}(\theta_{i})log(P_{\mathbb{B}}^{\Theta}(\theta_{i}))$}&$P_{\mathbb{B}}^{\Theta}(\theta_{i})=\frac{1}{|\Theta|}$&$\log (|\Theta|)$&\tabincell{l}{Elements;\\Cardinality;\\Mass function}\\
\hline
 Confusion measurement \cite{hohle1982entropy}& $C_{H}=-\sum_{F_{i}\in 2^{\Theta}}m(F_{i})\log Bel(F_{i})$&$m({\theta_{i}})=\frac{1}{|\Theta|}$&$\log (|\Theta|)$& \tabincell{l}{Mass function;\\Bel function} \\
 \hline
 Dissonance measurement \cite{yager1983entropy}&$E_Y=-\sum_{F_{i}\in 2^{\Theta}}m(F_{i})\log Pl(F_{i})$& \tabincell{l}{$m(F_i)=\frac{1}{K}, \forall 1 \rightarrow K,$\\$ \{F_1\}\cap \cdots \cap \{F_K\}= \varnothing $}& $\log (|\Theta|)$ &\tabincell{l}{Mass function;\\Pl function}\\
\hline
Hartley entropy  \cite{higashi1982measures}& $E_{H}=-\sum_{F_{i}\in 2^{\Theta}}m(F_{i})\log |F_{i}|$ &$m(\Theta)=1$&$\log (|\Theta|)$& \tabincell{l}{Mass function; \\Cardinality}  \\
\hline
Discord measurement \cite{klir1990uncertainty}& \tabincell{l}{$S_{KP}=-\sum_{F_{i}\in 2^{\Theta}}m(F_{i}) \log \sum_{G_{i}\in 2^{\Theta}}m(G_{i})\frac{|F_{i}\cap G_{i}|}{|G_{i}|}$}   &$m({\theta_{i}})=\frac{1}{|\Theta|}$&$\log (|\Theta|)$& \tabincell{l}{Mass function;\\ Cardinality} \\
\hline
 \tabincell{l}{Aggregate uncertainty \\(AU) measurement \cite{AU}}& $argmax_{\mathcal{P}}[-\sum_{i=1}^{n}p(\theta_{i})\log p(\theta_{i})]$&$BetP(\theta_{i})=\frac{1}{|\Theta|}$&$\log (|\Theta|)$& \tabincell{l}{Mass function}\\
\hline
\tabincell{l}{Pal \textit{et al.}'s entropy \cite{pal1993uncertainty}}&$E_{p}=-\sum_{F_{i}\in 2^{\Theta}} m(F_{i}) \log \frac{m(F_{i})}{{|F_{i}|}}$&$m(F_i)=\frac{|F_i|}{|\Theta|\cdot2^{|\Theta|-1}}$&$\log (|\Theta|\cdot 2^{|\Theta|-1})$&\tabincell{l}{Mass function;\\Cardinality}\\
\hline
Deng  entropy\cite{Deng2020ScienceChina}&$E_{d}=-\sum_{F_{i}\in 2^{\Theta}}m(F_{i})\log \frac{m(F_{i})}{2^{|F_{i}|}-1}$&\tabincell{l}{$m(A)=\frac{2^{|F_{i}|}-1}{\sum_{F_{i}\in 2^{\Theta}} 2^{|F_{i}|}-1}$}&$\log (3^{|\Theta|}-2^{|\Theta|})$&\tabincell{l}{Mass function;\\power set}\\
 \hline
 SU measurement \cite{wang2018uncertainty}& \tabincell{l}{$SU=\sum_{\theta_i\in\Theta}[-\frac{Pl(\theta_{i})+Bel(\theta_{i})}{2}\log\frac{Pl(\theta_{i})+Bel(\theta_{i})}{2}+Pl(\theta_{i})-Bel(\theta_{i})]$}& \tabincell{l}{$Bel(\theta_{i})=0; $\\$Pl(\theta_{i})=1$}&$|\Theta|$&\tabincell{l}{Bel function;\\Pl function;\\Elements}\\
 \hline
JS entropy \cite{jirouvsek2018new}&\tabincell{l}{$H_{JS}=\sum_{F_{i}\in 2^{\Theta}}m(F_{i})\log(|F_{i}|)-\sum_{i=1}^{n}Pl\_ P_{m}(\theta_{i})\log Pl\_ P_{m}(\theta_{i})$}&$m(\Theta)=1$&$2\log (|\Theta|)$&\tabincell{l}{PMT;\\Hartley entropy;\\Elements}\\
\hline
Decomposable entropy \cite{jirouvsek2020properties}&$H_q=\sum_{F_i\in 2^\Theta}(-1)^{|F_i|}q(F_i)\log q(F_i)$&NaN&NaN&\tabincell{l}{q function;\\Focal element}\\
\hline
Yang and Han's method \cite{yang2016new}&$TU^l=1-\frac{1}{n}\cdot\sqrt{3}\sum_{\theta_i\in\Theta}d^l([Bel(\theta_i),Pl(\theta_i)],[0,1])$&$m(\Theta)=1$&$1$&\tabincell{l}{Bel function;\\Pl function;\\Distnace}\\
 \Xhline{1.4pt}
\end{tabular}
\end{adjustbox}
\caption{These are classical and novel uncertainty measurements of BPA. We can find that in all of entropies, only JS entropy satisfies the intuitive maximum distribution.}
\label{d1t1}
\end{table*}
\restoregeometry

\end{definition}

\section{The process of probability transformation based on fractal} \label{process}
\subsection{Simulating probability transformation from fractal perspective}

\begin{figure}[htbp!]
\centering
\includegraphics[width=0.45\textwidth]{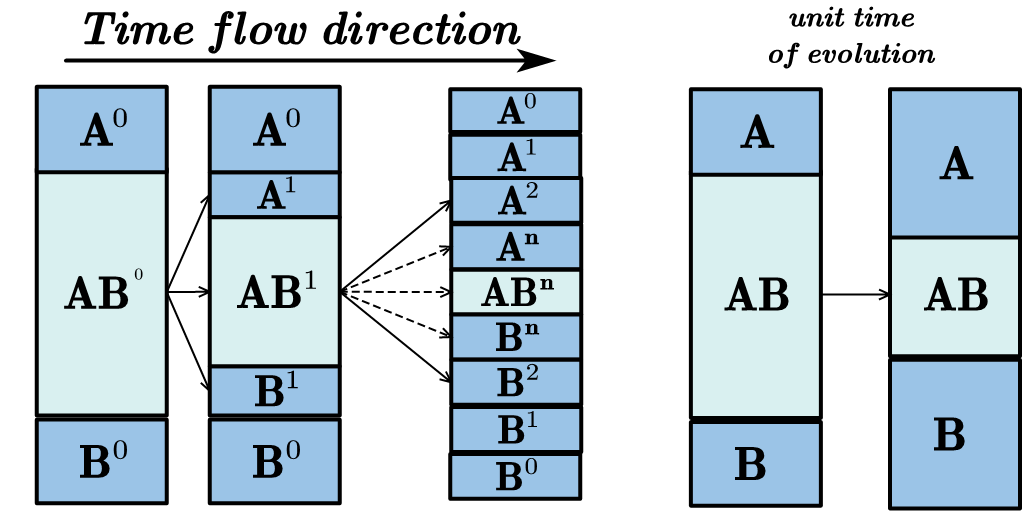}
\caption{The left part of Figure \ref{f1} shows that mass function of $AB$ is split to $A$ and $B$ in different ways over time. When $n\rightarrow \infty$, BPA transforms into probability distribution. For a certain unit time, right part is a step of transformation in a unit time, so we can think that the process of transformation is $AB$ continuously splitting itself into its own power sets $\{A, B, AB\}$.} 
\label{f1}
\end{figure}

Even though BPA can express more information by assigning the mass functions to the multi-element focal elements, in reality, all we observe are probability distributions. So how to reasonably transform BPA into probability distribution is the key to combining BPA with practical applications. PPT as the decision-making layer in TBM, which has wide applications. We propose a process for the PPT according to fractal, assuming that the result of PPT is generated under the action of time. For $2$-element discernment framework $X=\{A,B\}$, the process of BPA transforming into probability is shown in Figure \ref{f1}.

\begin{figure}[htbp!]
\centering
\includegraphics[width=0.45\textwidth]{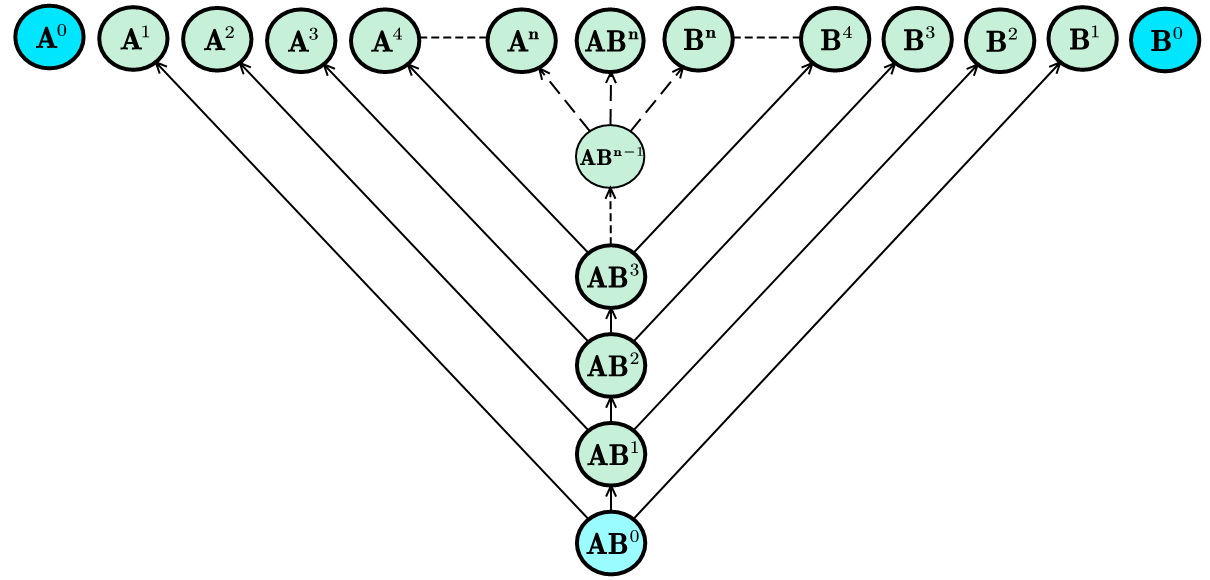}
\caption{The small triangle formed by the new split $AB^{n-1}$ and the large triangle of the entire split satisfy self-similarity.}
\label{f2}
\end{figure}

Self-similarity is a basic property of fractal theory, that is, in the process of fractal, the whole and part are similar.  In order to show this property more clearly, we use Figure \ref{f2} to show the process of splitting.
 
The entropy change in time due to the fractal geometry is assimilated to the information growth through the scale refinement. Wen \textit{et al.} also proved this point in their work on information dimension \cite{wen2021invited,Gao2021Information}. As shown in Figure \ref{f1} and \ref{f2}, as the number of splitting increases, before the newly generated $A$ and $B$ fusion the original $A$ and $B$, the overall belief entropy increases, which conforms to the ides proposed by Wang. The new BPA generated after the fusion means the system get new knowledge (the splitting method of original BPA), the overall belief entropy be unchanged or decreasing, which also conforms to information theory.
 
\subsection{The process of PPT}
 For a given BPA, when the probability transformation without receiving outside knowledge, the PPT  is most intuitive, and it uses an even splitting method to ensure the largest uncertainty of the information. According to the Equation \ref{PPTe} and the Figure \ref{f1}, with transformation in per unit time, it allocates equal mass functions to subsets with same cardinality, and the probability obtained at the end of the iteration must be PPT. Example \ref{ee1} shows the differences with different allocations.
 
 \begin{example}
 \label{ee1}
 Given a discernment framework $X=\{a,b,c\}$, and the BPA is $m(X)=1$. The change of mass belief function after per splitting are shown in follows,
\begin{equation}
\begin{aligned}
&m^{n}(F_{i})=m^{n-1}(F_{i})+\frac{1}{p}m^{n-1}(G_{i})+\frac{1}{q}m^{n-1}(H_{i})  \\
&m^{n}(G_{i})=(1-\frac{2}{p})m^{n-1}(G_{i})+\frac{1}{q}m^{n-1}(H_{i})\\
&m^{n}(H_{i})=(1-\frac{6}{q})m^{n-1}(H_{i}),
\end{aligned}
\end{equation}
where $p\geqslant 3$ and $q\geqslant 7$. $|F_{i}|=1,~|G_{i}|=2,~|H_{i}|=3$, and $m^{n}(A)$ means $n$ times splitting of $m(A)$. Because of $3$-element discernment framework has three $2$-element focal elements and one $3$-element focal element,  so $p$ and $q$ satisfy $p+4=q$.  When the $p$ and $q$ given different value, the change of mass functions with the splitting process is shown in the Figure \ref{nf1}. 
 \begin{figure}[htbp!]
\centering
\includegraphics[width=0.45\textwidth]{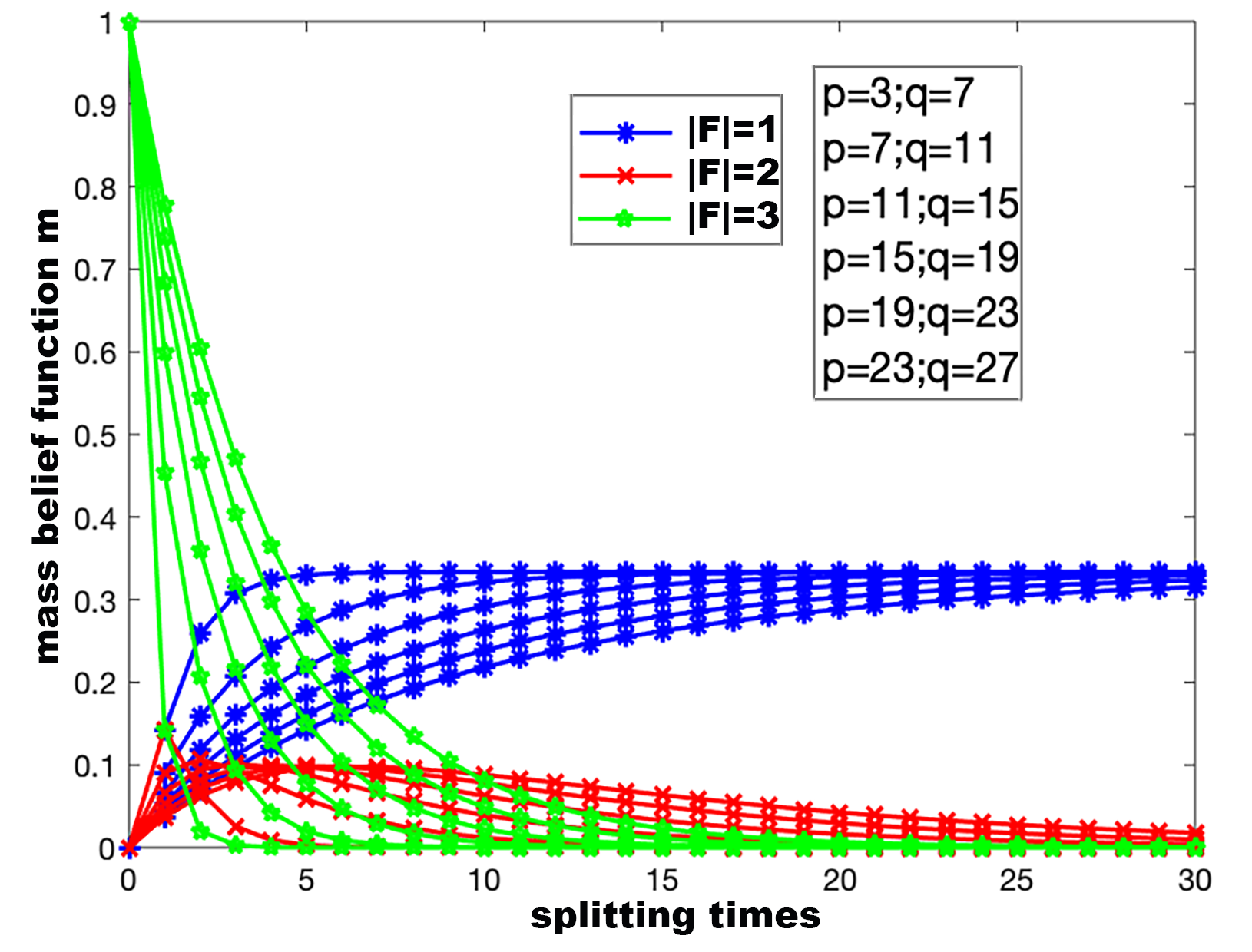}
\caption{Regardless of the values of $p$ and $q$, as the number of splits increases, BPA $m(X)=1$ is eventually transformed into a uniformly distributed probability $m(a)=m(b)=m(c)=\frac{1}{3}$, which is the same as the result of PPT.}
\label{nf1}
\end{figure}

Hartley entropy \cite{higashi1982measures} represents the uncertainty of non-specificity in BPA. When Hartley entropy is $0$, BPA degenerates into probability distribution. So as shown in Figure \ref{fh1}, in the process of BPA transformation into probability distribution, Hartley entropy of BPA gradually decreases from the maximum value to $0$.

 \begin{figure}[hbtp!]
\centering
\includegraphics[width=0.95\textwidth]{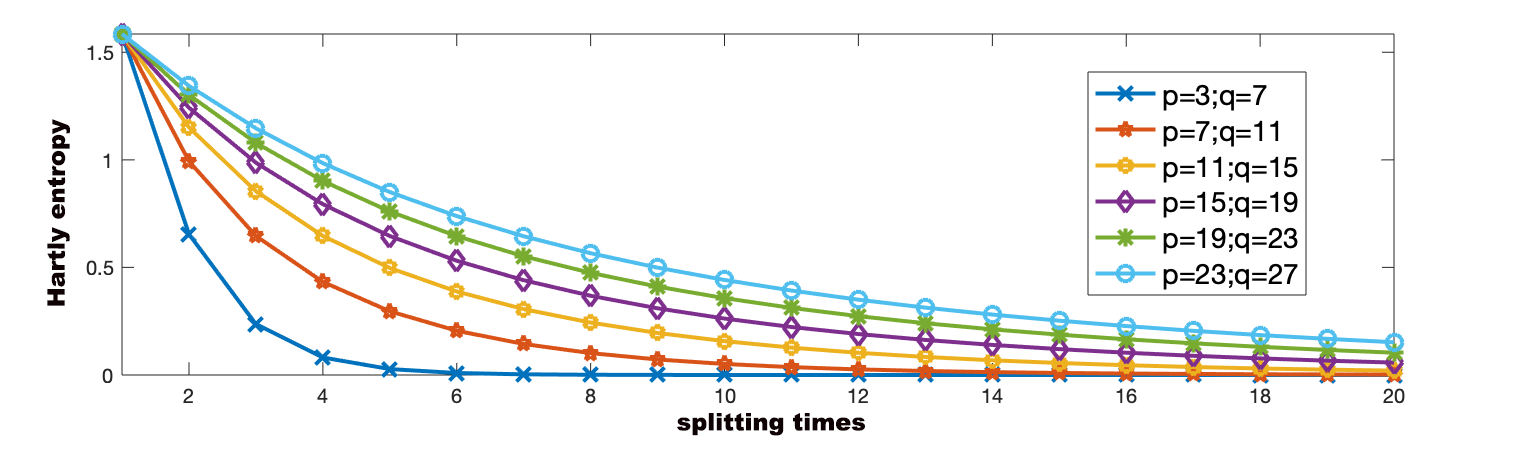}
\caption{The trend of Hartley entropy in Example \ref{ee1}.}
\label{fh1}
\end{figure}
 \end{example}
 
According to the above description, the PPT process is shown in detail, and the transformation process is different for different scales of unit time division. But the result is to maintain an uniform distribution for all elements.

\subsection{Discussion the process of probability transformation}

Besides the PPT, other methods also can be written as the fractal process if they satisfy the upper and low probability requirement. But other methods only give the calculation methods of results, and result-oriented inference cannot accurately simulate the transformation process, so we won't discuss them here. Although PMT cannot be written in the form of a split mass function, it can be seen as a continuous fusion of a uniform focal element assignment $m(F_i)=\frac{1}{2^{|F_i|-1}}$. Example \ref{eppf} shows that the transformation process of PMT in Definition \ref{PPF}.

\begin{example}\label{eppf}
 Given a discernment framework $X=\{a,b\}$, and its BPA is $\mathbb{B}(2^{\Theta})=\{m(a)=0.2,m(b)=0.4,m(ab)=0.4\}$. Based on Definition \ref{PPF}, the PMT of $\mathbb{B}(2^{\Theta})$ is \\ $\{PnPl(a)=\frac{3}{7},PnPl(b)=\frac{4}{7}\}$. If we continually use another BPA $\mathbb{B}(2^{X})=\{m(a)=\frac{1}{3},m(b)=\frac{1}{3},m(ab)=\frac{1}{3}\}$ to fuse the $\mathbb{B}(2^{\Theta})$ by Dempster combination rule, the results are shown in Figure \ref{fppf}.
 
\begin{figure}[htbp!]
\centering
\includegraphics[width=0.45\textwidth]{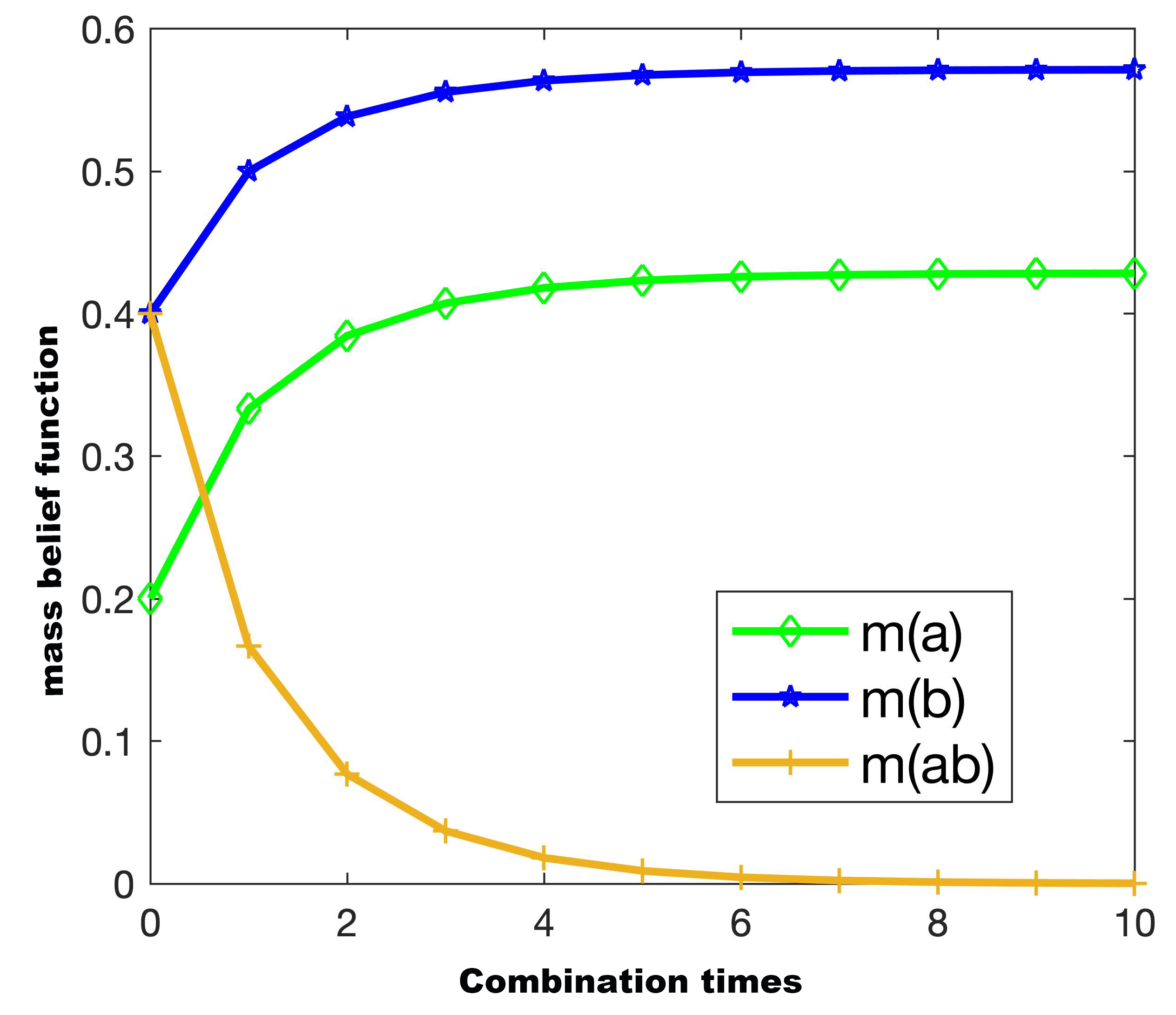}
\caption{With the continuous fusion $\mathbb{B}(2^{X})$ in Example \ref{eppf}, $\mathbb{B}(2^{\Theta})$ eventually transform into $PnPl$.}
\label{fppf}
\end{figure}

\end{example}

In this section, we present the implementation process of the existing main probability transformation methods. For the newly proposed probability transformation methods, the rationality can be verified according to the process ideas given in this section. More importantly, for BPA, its uncertainty can be measured by using the intermediate quantity of its transformation process. The specific method will be given in next section.

\section{Fractal-based belief entropy}\label{fbentropy1}

A new belief entropy called fractal-based belief (FB) entropy is proposed based on the process of PPT in this section. It can not only measure the uncertainty of BPAs, but also make their maximum entropy distributions correspond to solving actual physical problems. In Example \ref{ee1}, when $p$ and $q$ take different values, the evolution speed of PPT in per unit time is different. In order to better express the concept of "uniformity", we rule the focal element is equally split into its power set in per unit time.

\subsection{Fractal-based belief entropy}
\begin{definition}[FB entropy]\label{bfentropy}
For a discernment framework $\Theta=\{\theta_{1},\theta_{2},\dots,\theta_{n}\}$, its BPA is $\mathbb{B}(2^{\Theta})$, and the fractal-based (FB) entropy of $\mathbb{B}(2^{\Theta})$ is defined as
\begin{equation}\label{bfentropye}
%E_{BF}(\Theta)=-\sum_{A\subseteq X}[(\frac{m(A)}{2^{|A|}-1}+\sum_{A\subseteq B~and~|A|<|B|}\frac{m(B)}{2^{|B|}-1})\log(\frac{m(A)}{2^{|A|}-1}+\sum_{A\subseteq B~and~|A|<|B|}\frac{m(B)}{2^{|B|}-1}) ]
E_{FB}=-\sum_{F_{i}\subseteq \Theta(F_{i}\in 2^{\Theta})} m_{F}(F_{i})\log m_{F}(F_{i}),
\end{equation}
where
\begin{equation}\label{mfe}
m_{F}(F_{i})=\frac{m(F_{i})}{2^{|F_{i}|}-1}+\sum_{F_{i}\subseteq G_{i}\cap|F_{i}|<|G_{i}|}\frac{m(G_{i})}{2^{|G_{i}|}-1}.
\end{equation}
The new set $\mathbb{B}_{F}(2^{\Theta})$ composed by $m_{F}(F_{i})$ is called fractal-based basic probability assignment (FBBPA).
\end{definition}

By observing the Equation \ref{mfe}, we can find that $\mathbb{B}_{F}(2^{\Theta})$ is obtained by a unit time transformation of PPT. For $m(\Theta)=1$, the most uncertain BPA intuitively, after a unit time splitting, the  $\mathbb{B}_{F}(2^{\Theta})$ is a uniform distribution of $2^{\Theta}$, which is same with the maximum entropy distribution of Shannon entropy.  So FBBPA  is neither BPA nor probability distribution, but describes the characteristics of BPA from the perspective of probability.

\subsection{The Maximum FB entropy and its physical meaning}
\begin{definition}[Maximum FB entropy]\label{maxbfentropy}
For a discernment framework $\Theta=\{\theta_{1},\dots,\theta_{n},\}$, its BPA is $\mathbb{B}(2^{\Theta})$. The maximum fractal-based belief entropy $E_{FB}^{\uparrow}(\Theta)$ is
\begin{equation}\label{maxbfentropy}
\begin{aligned}
E_{FB}^{\uparrow}=\log (2^{n}-1),
\end{aligned}
\end{equation}
when $m(\Theta)=1.$
\end{definition}

\textbf{\emph{Proof.}}Let
\begin{equation}
E=-\sum_{F_{i}\in2^\Theta}m_{F}(F_{i})\log m_{F}(F_{i}),
\end{equation}
and according to Equation \ref{mfe}, it is obvious that $\sum_{A\subseteq \Theta}m_{F}(A)=1$. So the Lagrange function can be denoted as
\begin{equation}
E_{0}=-\sum_{F_{i}\in2^\Theta}m_{F}(F_{i})\log m_{F}(F_{i})+\lambda (\sum_{F_{i}\subseteq \Theta}m_{F}(F_{i})-1),
\end{equation}
and calculate its gradient
\begin{equation}
\frac{\partial E_{0}}{\partial m_{F}(F_{i})}=-\log m_{F}(F_{i})-\frac{1}{\ln a}+\lambda=0.
\end{equation}
For all $F_{i}\subseteq \Theta$
\begin{equation}
 \log m_{F}(F_{i})=-\frac{1}{\ln a}+\lambda=k,
 \end{equation}
so when $m_{F}(F_{i})=\frac{1}{2^{|\Theta|}-1}$ and $m(\Theta)=1$, the $E_{FB}(\Theta)$ reaches the maximum.

The maximum Shannon entropy called information volume and its probability distribution can solve real physical problems in practical applications. As the generalization of the Shannon entropy, the maximum FB entropy also has a corresponding physical model in reality, which are shown in Example \ref{e2}.
\begin{example}[Physical model of maximum FB entropy]\label{e2}
Assuming there are $64$ teams participating in a competition. The only information source is organizer, and we can ask him whether some teams are champions. The goal for us is to find all champions.
\begin{description}
\item[\textbf{Q}:] How many times inquiring can we find the champion at least?
\item[\textbf{Case1}:]We know the number of champions is $1$.
\item[\textbf{Case2}:]We don’t know the exact number of champions.
\end{description}
Figure \ref{FB_PHYSICAL} shows the difference between the t $2$ Cases.
\begin{figure}[htbp!]
\centering
\includegraphics[width=0.8\textwidth]{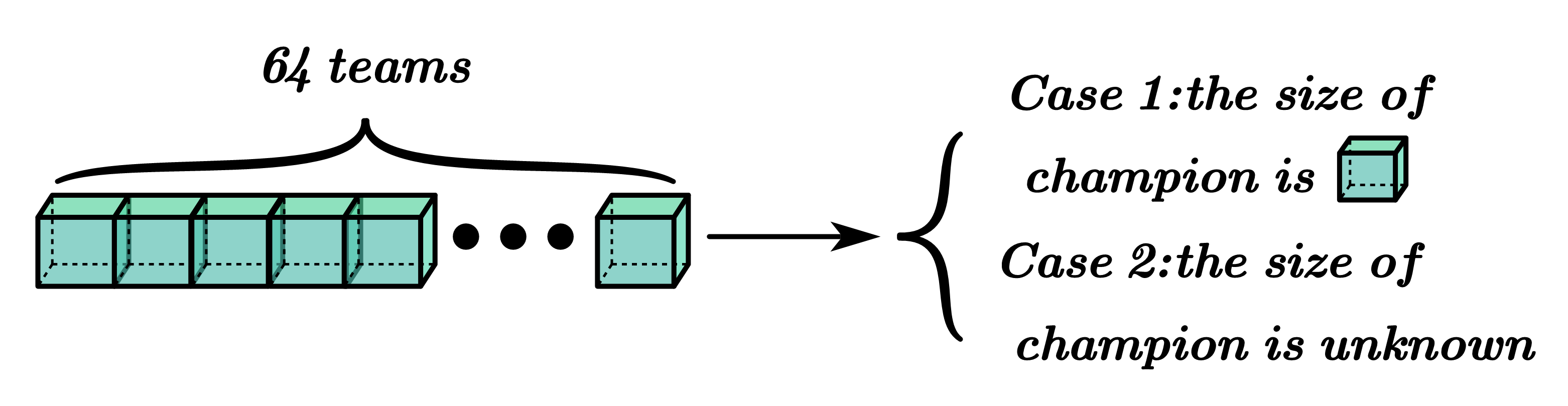}
\caption{The difference between $2$ Cases.}
\label{FB_PHYSICAL}
\end{figure}

Case $1$ can be written as an uniform probability distribution with $64$ basic events $\{p(1)=\cdots=p(64)=\frac{1}{64}\}$. The Shannon entropy with base $2$ is $\log_2 64=6$, so only $6$ times inquiring can we find the champion. But for Case $2$, we are not sure about the number of championships, so all power sets of the $64$ teams frame have equal probability to win championships. It can be written as $\{p(1)=\frac{1}{2^{64}-1},p(2)=\frac{1}{2^{64}-1},\cdots,p(1\cdots64)=\frac{1}{2^{64}-1}\}$, which also corresponds to the FBBPA of the maximum FB entropy $\{m_F(1)=\frac{1}{2^{64}-1},m_F(2)=\frac{1}{2^{64}-1},\cdots,m_F(1\cdots64)=\frac{1}{2^{64}-1}\}$, so it corresponds to BPA $m(1\cdots64)=1$ and FB entropy $E_{FB}=\log_2 (2^{64}-1)\approx64$. The inquiry times of Case $2$ is $64$, which means that we can only find all the champions by inquiring all teams. 

\end{example}

Example \ref{e2} illustrates that FB entropy is a generalization of Shannon entropy in the physical model of maximum entropy.

\subsection{Evaluation FB entropy}

According to the $10$ requirements for total uncertainty measurements of BPA in \cite{klir2013uncertainty,abellan2008requirements}, we evaluate the properties of FB entropy to prove its advantages.  Among them, $\heartsuit$ means that FB entropy satisfies this proposition, and $\spadesuit$ means that FB entropy does not satisfy this proposition. For the unsatisfied propositions, we give explanations and prove the rationality of FB entropy. Under an $n$-element discernment framework $\Theta$ with BPA $\mathbb{B}(2^{\Theta})$, we evaluate the FB entropy in Proposition \ref{p1} - \ref{p10}.

 \begin{proposition}[Probabilistic consistency ($\heartsuit$)]\label{p1}
When $\forall |F_i|>1$ $m(F_i)=0$, the total uncertainty measurement should degenerate into the Shannon entropy.
  \end{proposition}

  \textbf{\emph{Proof.}}When $\mathbb{B}(2^{\Theta})$ satisfies $\sum_{\theta_i\in\Theta}m(\theta_i)=1$, substitute it into Equation \ref{bfentropye} and \ref{mfe}:
  \begin{equation}
  E_{FB}=-\sum_{F_{i} \in 2^\Theta} m_{F}(F_{i})\log m_{F}(F_{i}) = -\sum_{i=1}^{n}m(\theta_{i})\log m(\theta_{i})=H(\mathbb{B}).
   \end{equation}
  So the FB entropy satisfies the Proposition \ref{p1}.  
  
   \qed

 \begin{proposition}[Set consistency($\spadesuit$)]\label{p2}
The total uncertainty measurement of vacuous BPA ($m(\Theta)=1$) should equal to Hartley entropy $TU=E_H=\log |\Theta|$.
 \end{proposition}

 \textbf{\emph{Proof.}} For vacuous BPA, $E_{FB}=\log (2^{|\Theta|}-1) \neq \log |\Theta|\neq E_H.$

 So the FB entropy doesn't satisfy the Proposition \ref{p2}.

\begin{figure}
  \begin{minipage}[htbp]{0.5\linewidth}
    \centering
    \includegraphics[width=0.98\textwidth]{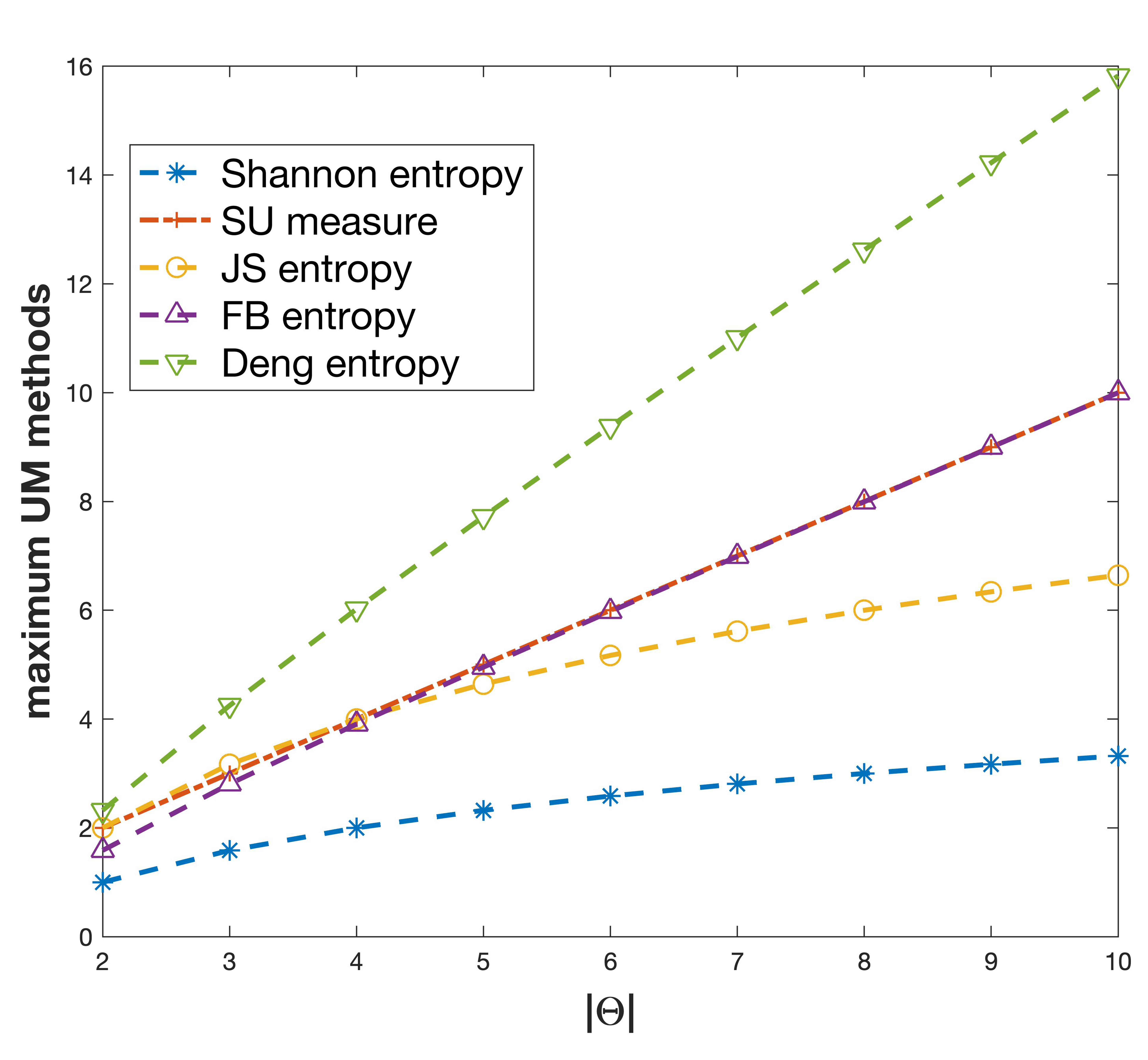}
    \caption{The maximum uncertainty of common total uncertainty measurements are larger than maximum Shannon entropy, so the requirement of set consistency is not reasonable.}
    \label{f5}
  \end{minipage}
  \begin{minipage}[htbp]{0.5\linewidth}
    \centering
    \includegraphics[width=0.98\textwidth]{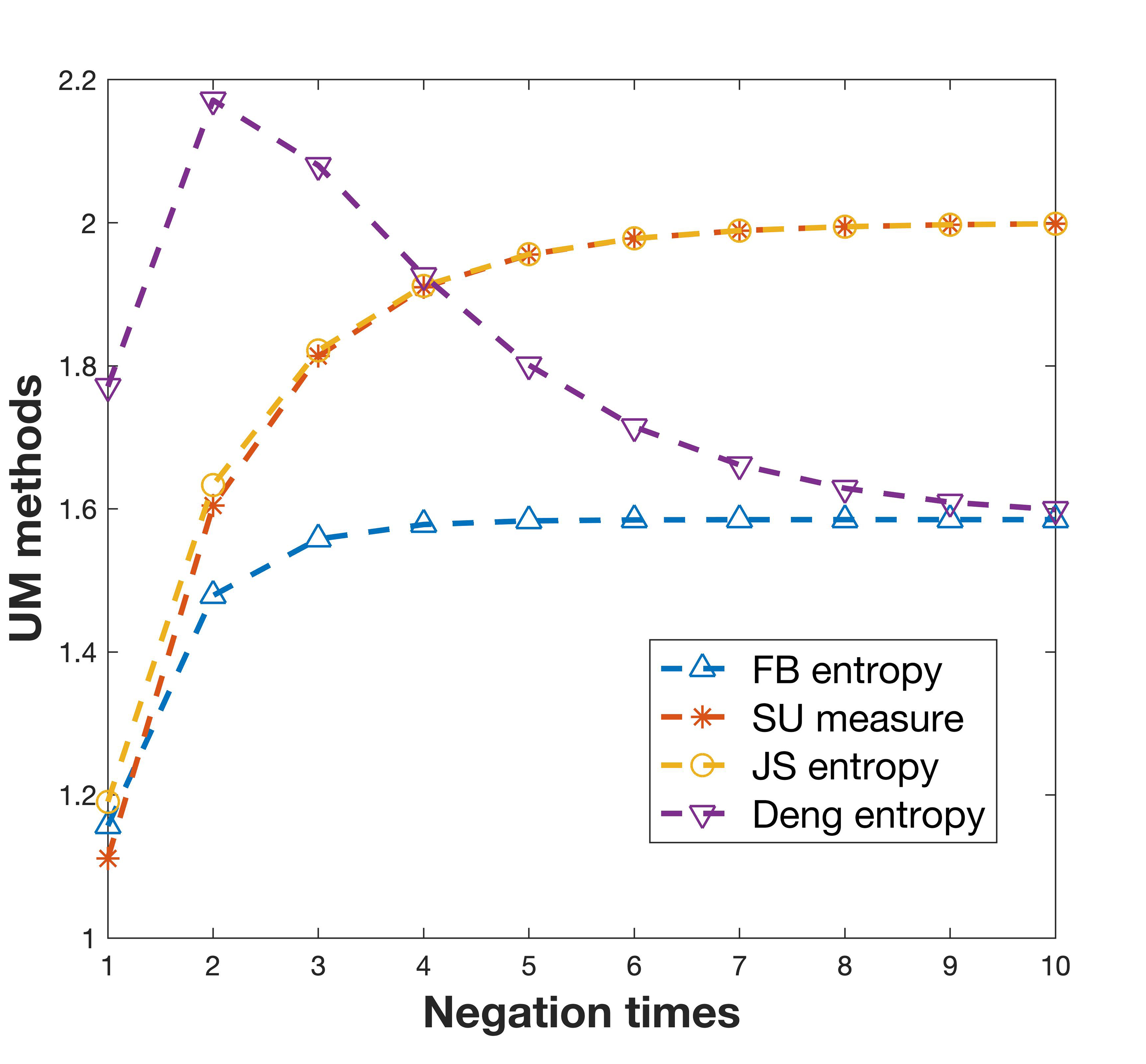}
    \caption{With the uncertainty increasing, the FB entropy, JS entropy and SU measurement is also increasing.}
    \label{f6}
  \end{minipage}
\end{figure}

\textbf{\emph{Explanation.}}The uncertainty of probability distribution is caused by the discord between basic events. The maximum Shannon entropy probability distribution is uniform, which means assignment same support degree to all events. In DSET, BPA can not only reflect the discord between elements, but contain the uncertainty to the distribution itself as well. In Example \ref{e2}, a BPA under $n$-element discernment framework and a probability distribution under $2^n-1$-events random variables can express same information, which also express that BPA can express more information than probability distribution for same dimension. So the maximum belief entropy larger than maximum Shannon entropy is more rational. Some common maximum total uncertainty measurement and Shannon entropy are shown in Figure \ref{f5} to show this property more intuitively.

\qed

 \begin{proposition}[Monotonicity($\heartsuit$)]\label{p6}
If BPA $\mathbb{B}(2^{\Theta})$ and $\mathbb{B}(2^{\Omega})$ have following relationship: $\mathbb{B}(2^{\Theta}) \subseteq \mathbb{B}(2^{\Omega})$, the total uncertainty measurements  of them should satisfy $UM(\mathbb{B}(2^{\Theta})) \leq UM(\mathbb{B}(2^{\Omega}))$.
 \end{proposition}

 \textbf{\emph{Proof.}}
 Luo \textit{et al.} \cite{Luo2019matrix} proposes a widely used method of  BPA's negation, and the direction of negation is the direction of ignorance. For a discernment framework $\Theta=\{\theta_{1},\theta_{2}\}$, its process of negation in $10$ times is shown in Table \ref{t2}, and the trends of JS entropy, SU measurement, Deng entropy and FB entropy are shown in Figure \ref{f6}. According to their trends, we can find that  JS entropy, SU measurement and FB entropy is continuously rising, which illustrate that they satisfy the Proposition \ref{p6}. But for Deng entropy, its maximum entropy distribution is a uniform distribution based on time, so it does not satisfy monotonicity in this case.

\begin{table}[htbp!]
\caption{Intuitively, as the number of negation increases, $m(\Theta)$ gradually increases, and the uncertainty expressed by BPA be larger}
\label{t2}
\begin{center}
\begin{tabular}{c|cccccccccc}
  \Xhline{1.4pt}
  Times& $1$& $2$& $3$& $4$& $5$& $6$& $7$& $8$& $9$& $10$ \\
 \Xhline{1.4pt}
$m(x_{1}) $       & $0.6000$& $0.0500$& $0.1500$& $0.0125$& $0.0375$&  $0.0031$& $0.0094$& $0.0008$& $0.0023$& $0.0002$\\
$m(x_{2})$        & $0.1000$& $0.3000$& $0.0250$& $0.0750$& $0.0063$ &  $0.0187$& $0.0016$& $0.0047$& $0.0004$& $0.0012$ \\
$m(x_{1}x_{2})$& $0.3000$& $0.6500$& $0.8250$& $0.9125$& $0.9562$ & $0.9781$& $0.9891$& $0.9945$& $0.9973$& $0.9986$  \\
  \Xhline{1.4pt}
\end{tabular}
\end{center}
\end{table}

\qed

 \begin{proposition}[Range($\spadesuit$)]\label{p3}
 The range of total uncertainty measurements should satisfy the $[0,\log (|\Theta|)]$
 \end{proposition}

 \textbf{\emph{Proof.}}
 If $m(\theta_{i})=1$, the FB entropy reaches the minimum $0$. So the $E_{FB}^{\downarrow}=0$. In the Definition \ref{maxbfentropy}, we have proven that the $E_{FB}^{\uparrow}=\log(2^{n}-1)$. Based above, the range of FB entropy is $[0,\log (2^n-1)]$, which doesn't satisfy the Proposition \ref{p3}. 

The explanation of Proposition \ref{p3} is similar to Proposition \ref{p2}.

\qed

 \begin{proposition}[Additivity($\heartsuit$)]\label{p4}
Suppose $X$, $Y$ and $Z$ are $3$ discernment frameworks. Among them, $X$ and $Y$ are independent, and $Z=X\times Y$. The total uncertainty measurement should satisfy
\begin{equation}
TUM(Z)=TUM(X)+TUM(Y),
\end{equation}
where $TUM$ is a general term for total uncertainty measurements.
\end{proposition}

\begin{figure}[htbp!]
\centering
\includegraphics[width=0.85\textwidth]{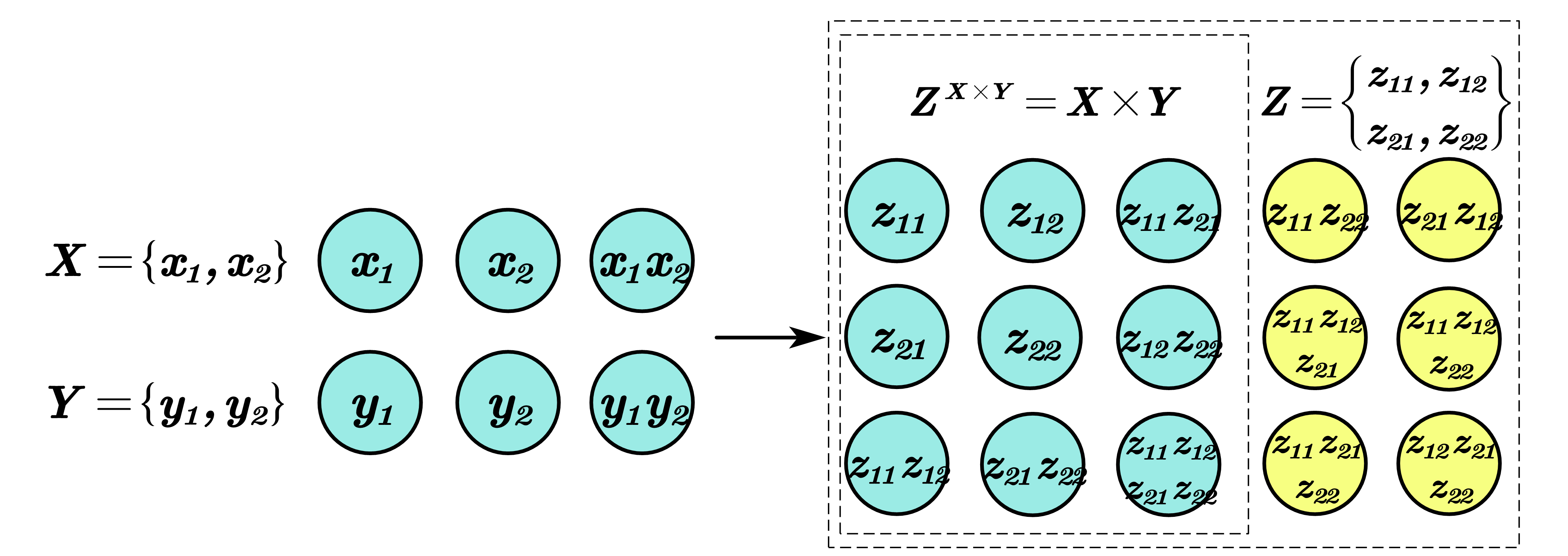}
\caption{Suppose $2$ discernment frameworks $X=\{x_1,x_2\}$ and $Y=\{y_1,y_2\}$, and we use the Smets' joint frame can get a new frame $Z=\{z_{11},z_{12},z_{21},z_{22}\}$. For the normalized BPA, original BPAs' product can not cover all focal elements under new frame, so the mass functions of yellow parts are always be $0$, which appears conflict with the definition of joint probability distribution. So we define the mass functions of joint BPA are generated by the product of original BPAs (blue part).}
\label{f61}
\end{figure}

\textbf{\emph{Proof.}} Joint BPA has different definitions according to whether $m(\varnothing)=0$. Smets \cite{smets1993belief} defined that generalized Bayesian theorem under the condition $m(\varnothing)\neq 0$ and for the joint frame $\Psi=\Theta \times \Omega$, the number of mass functions satisfies $2^{\Psi}=2^{\Theta + \Omega}$. But in this paper, we consider BPA is normalized, so the number of joint mass functions for $X \times Y$ is $(2^{|X|}-1)(2^{|Y|}-1)$. Figure \ref{f61} using a specific case to show the difference between $2$ definitions intuitively.

According to the above description, the number of mass functions of the joint BPA is less than the power set under the joint framework. According to Definition \ref{bfentropy}, the calculation method of FBBPA at this time is no longer assigning mass functions to its power set, but assigning them to subsets which has inclusive relationship under current frame.
For joint frame $Z =X\times Y $, we define joint BPA $m^{Z}$ and joint FBBPA  $m_{F}^Z$ as follows: 

\begin{equation}\label{mze}
\begin{aligned}
&m^Z(z_{ij})=m(x_i)\times m(y_j);\\
&m^Z(z_{ij}z_{im})=m(x_i)\times m(y_j y_m);\\
&m^Z(z_{ij}z_{im}z_{nj}z_{nm})=m(x_i x_n) \times m(y_j y_m);
\end{aligned}
\end{equation}

\begin{equation}\label{mzfe} 
\begin{aligned}
\forall F_i \subseteq Z,m_{F}^Z(F_i)=m^Z(F_i)+\sum_{F_i\subseteq K_i;G_i\times H_i =K_i}\frac{m^Z(K_i)}{(2^|G_i|-1)(2^|H_i|-1)};
\end{aligned}
\end{equation}

For $C\subseteq Z$, $A\subseteq X$ and $B \subseteq Y$ and they satisfy $A\times B=C$, according to the Equation \ref{mze} and \ref{mzfe},

\begin{equation}
\begin{aligned}
m^{Z}_F(C)&=m^Z(C)+\sum_{C \subseteq K_i;G_i\times H_i =K_i;A\subseteq G_i;B\subseteq H_i }\frac{m^Z(K_i)}{(2^|G_i|-1)(2^|H_i|-1)}\\
&=m(A)\times m(B)+\sum_{C \subseteq K_i;G_i\times H_i =K_i ;A\subseteq G_i;B\subseteq H_i }\frac{m(G_i)\times m(H_i)}{(2^|G_i|-1)(2^|H_i|-1)}\\
&=(m(A)+\sum_{A\subseteq G_i}\frac{m(G_i)}{2^|G_i|-1})\times(m(B)+\sum_{B\subseteq H_i}\frac{m(H_i)}{2^|H_i|-1})\\
&=m_F(A)\times m_F(B)
\end{aligned}
\end{equation}

We know that the Shannon entropy satisfies additivity, and it is proved that the consistency of the joint FBBPA and the joint BPA, it is easy to conclude that the FB entropy satisfies the additivity and Example \ref{eee} shows its calculation process. So the FB entropy satisfies the Proposition \ref{p4}.

\begin{example}\label{eee}
For two independent BPAs under $2$-element frames $X$ and $Y$, $\mathbb{B}(2^X)=\{m(x_1)=m(x_2)=\frac{1}{5},m(x_1 x_2)=\frac{3}{5}\}$ and $\mathbb{B}(2^Y)=\{m(y_1)=\frac{1}{10},m(y_2)=\frac{3}{5},m(y_1 y_2)=\frac{3}{10}\}$, so the joint BPA is
\begin{equation}
\begin{aligned}
 &\mathbb{B}(2^Z)=\mathbb{B}(2^X)\times \mathbb{B}(2^Y)=\{m(z_{11})=\frac{1}{50},m(z_{12})=\frac{6}{50},m(z_{21})=\frac{1}{50},m(z_{22})=\frac{6}{50},\\
 &m(z_{11} z_{21})=\frac{3}{50},m(z_{12} z_{22})=\frac{18}{50},m(z_{11} z_{12})=\frac{3}{50},m(z_{21} z_{22})=\frac{3}{50},m(z_{11} z_{12} z_{21} z_{22})=\frac{3}{50}\}.
 \end{aligned}
 \end{equation}
 According to Equation \ref{mzfe}, the joint FBBPA is
 
 \begin{equation}
\begin{aligned}
 &\mathbb{B}_F(2^Z)=\mathbb{B}_F(2^X)\times \mathbb{B}_F(2^Y)=\{m_F(z_{11})=\frac{4}{50},m_F(z_{12})=\frac{14}{50},m_F(z_{21})=\frac{4}{50},m_F(z_{22})=\frac{14}{50},\\
 &m_F(z_{11} z_{21})=\frac{2}{50},m_F(z_{12} z_{22})=\frac{7}{50},m_F(z_{11} z_{12})=\frac{2}{50},m_F(z_{21} z_{22})=\frac{2}{50},m_F(z_{11} z_{12} z_{21} z_{22})=\frac{1}{50}\}.
 \end{aligned}
 \end{equation}
 So the FB entropy of $\mathbb{B}(2^Z)$, $\mathbb{B}(2^X)$ and $\mathbb{B}(2^Y)$ satisfy that $E_{FB}(Z)=2.6787=1.5219+1.1568=E_{FB}(X)+E_{FB}(Y)$.
 
\end{example}
 
\qed

 \begin{proposition}[Subadditivity($\heartsuit$)]\label{p5}
Suppose $X$, $Y$ and $Z$ are discernment frameworks. And $Z=X\times Y$. The total uncertainty measurements should satisfy
\begin{equation}
TUM(Z) \leq TUM(X)+TUM(Y).
\end{equation}
 \end{proposition}

 \textbf{\emph{Proof.}}
\begin{description}
\item[\textbf{Case1:}] If the BPAs of $X$ and $Y$ are independent, the $E_{FB}(Z) = E_{FB}(X)+E_{FB}(Y)$ has been proven in Propositon \ref{p4}.
\item[\textbf{Case2:}] If the BPAs of $X$ and $Y$ are not independent, when they are combined into a joint BPA, they obtain information from each other's BPA, which can reduce the uncertainty of the joint BPA. So $E_{FB}(Z) < E_{FB}(X)+E_{FB}(Y)$ .
\end{description}
 According to $2$ Cases, we can prove that the FB entropy satisfies the Proposition \ref{p5}.

\qed

 \begin{proposition}[RB1($\heartsuit$)]\label{p7}
The calculation process of total uncertainty measurement cannot be too complicated.
  \end{proposition}

 \textbf{\emph{Proof.}}
For the $n$-element discernment framework, the computation complexity of Equation \ref{bfentropye} and \ref{mfe} are $\mathcal{O}(2^n)$ and $\mathcal{O}(n2^n)$ respectively. According to Table \ref{d1t1}, the computation complexity of JS entropy and SU is similar with FB entropy, and Yang and Han's method has higher complexity than them. Therefore, the complexity of FB entropy is within an absolutely acceptable range and satisfies Proposition \ref{p7}.
 
 \qed
 
 \begin{proposition}[RB2($\heartsuit$)]\label{p8}
The total uncertainty measurement can be divided two methods to measure the discord and non-specificity respectively.
\end{proposition}
  \textbf{\emph{Proof.}}

Different from other methods (JS entropy, SU measurement and Deng entropy), the discord and non-specificity can not be divided from the expression directly, but it can obtain these two measurements in a more reasonable way. We define FB entropy from the process of PPT, so when BPA transformed to PPT, the assignment expresses discord only. Based on above, the discord $E^{\mathcal{D}}_{FB}$ and non-specificity $E^{\mathcal{N}}_{FB}$ of $E_{FB}$ are defined as follows:
\begin{equation}
\begin{aligned}
&E^{\mathcal{D}}_{FB}=H(BetP)=H_j=-\sum_{\theta_i\in\Theta}BetP(\theta_{i})\log BetP(\theta_{i});\\
&E^{\mathcal{N}}_{FB}=E_{FB}-E^{\mathcal{D}}_{FB}=\sum_{\theta_i\in\Theta}BetP(\theta_{i})\log BetP(\theta_{i})-\sum_{F_i\in2^\Theta} m_{F}(F_i)\log m_{F}(F_i).
\end{aligned}
\end{equation}

\begin{figure}[htbp!]
\centering
\includegraphics[width=0.75\textwidth]{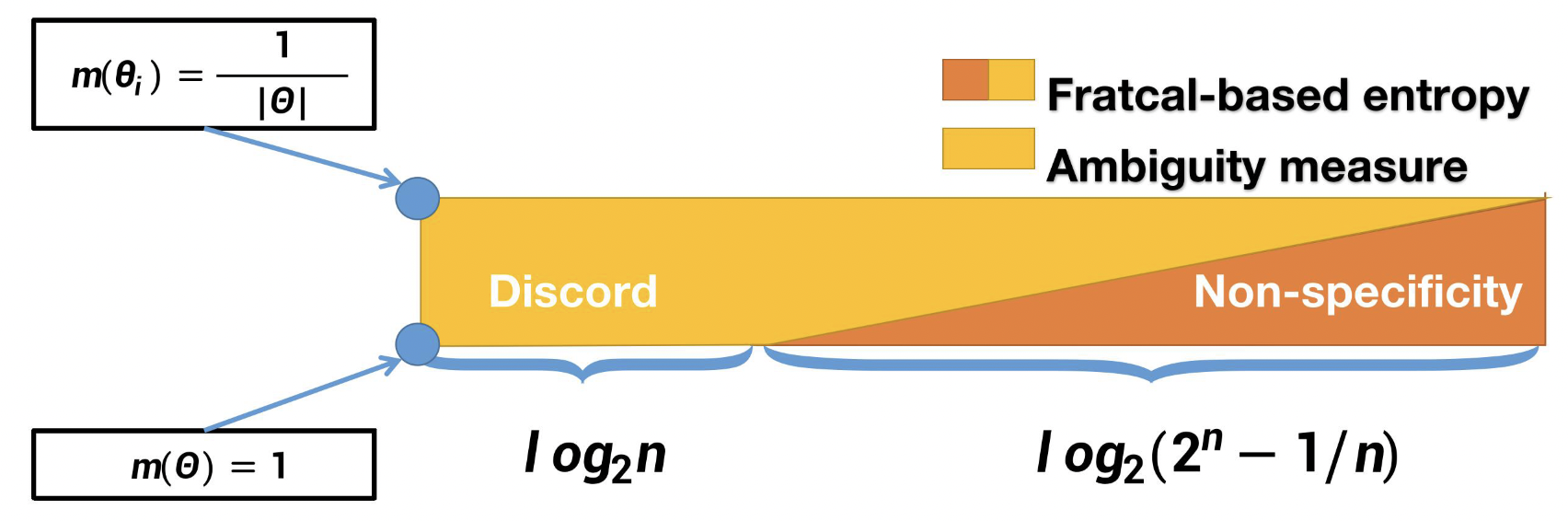}
\caption{The relationship of FB entropy and its discord $\&$ non-specificity measure}
\label{f71}
\end{figure}

The relationship of discord and non-specificity are shown in Figure \ref{f71}. Based on above, the FB entropy satisfies Proposition \ref{p8}.

 \begin{proposition}[RB3($\heartsuit$)]\label{p9}
Total uncertainty measurement must be sensitive to changes in BPA.
\end{proposition}

  \qed

  \textbf{\emph{Proof.}}
Since the change from BPA to FBBPA is reversible, any change to BPA equals to change FBBPA. For different FBBPA, Shannon entropy has been proved to be a sensitive measurement method, so for any BPA, FB entropy is also sensitive to its changes. So FB entropy satisfies Proposition \ref{p9}. We use Example \ref{e3} to show the results intuitively.
 
 \qed
 
  \begin{example}\label{e3}
  For discernment framework $X=\{x_1,x_2\}$, $m(x_1)$ and $m(x_2)$ change from $0$ to $1$ and satisfy $m(x_1)+m(x_2)\leqslant 1$. 
  Figure \ref{f74} and \ref{f75} are the change trend of discord and non-specificity, and Figure \ref{f72} and \ref{f73} show their relationships and top view. 
  
    \begin{figure}
  \begin{minipage}[htbp]{0.48\linewidth}
    \centering
    \includegraphics[width=0.98\textwidth]{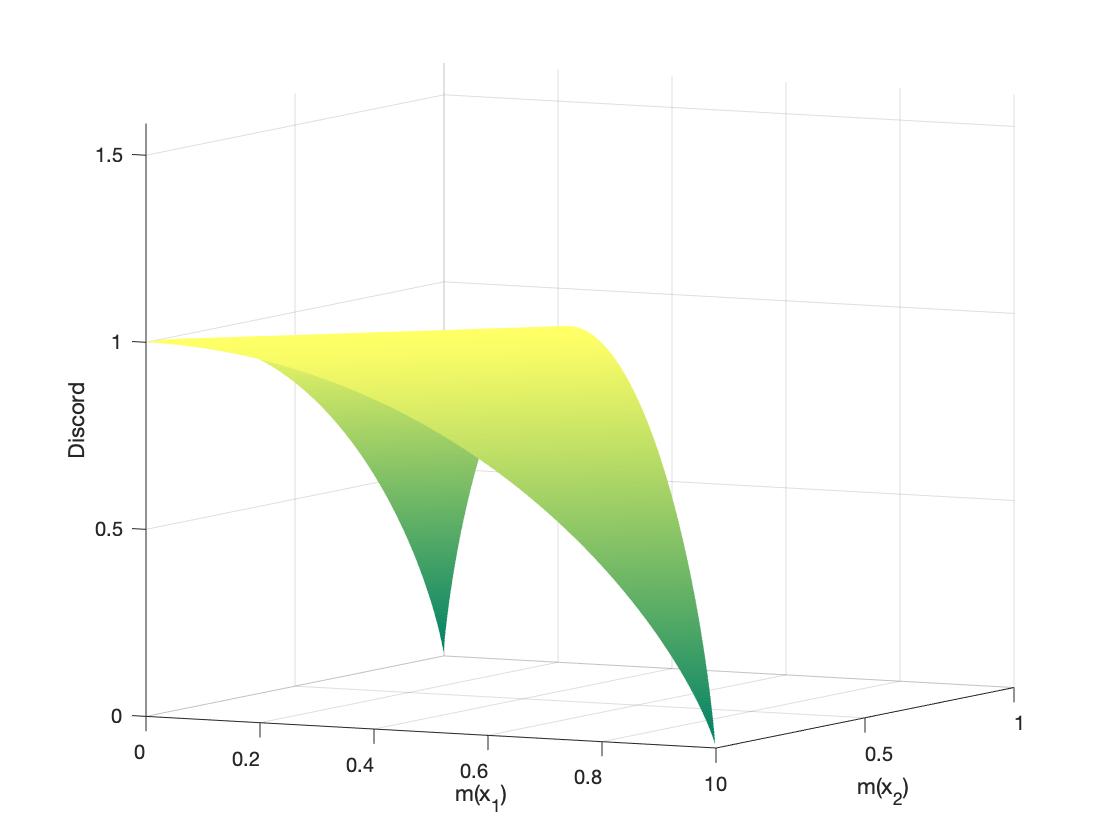}
    \caption{Discord change trend in Example \ref{e3}}
    \label{f74}
  \end{minipage}
  \begin{minipage}[htbp]{0.48\linewidth}
    \centering
    \includegraphics[width=0.98\textwidth]{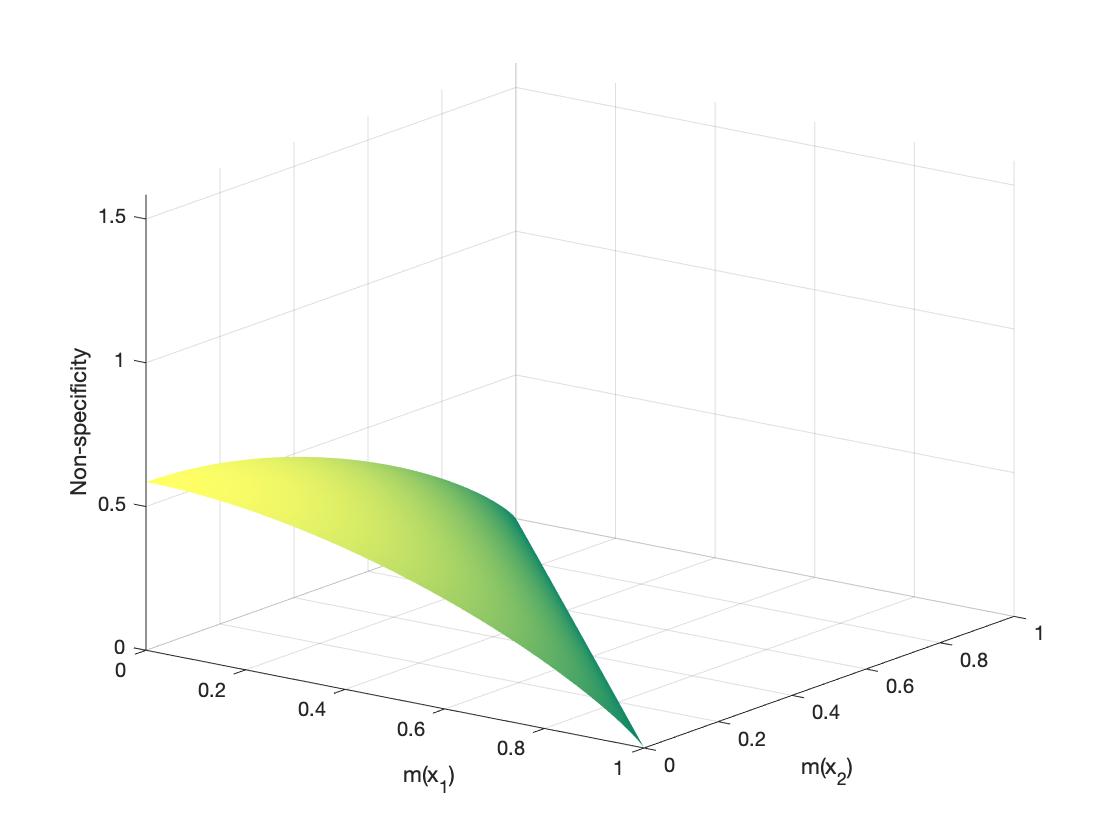}
    \caption{Non-specificity change trend in Example \ref{e3}}
    \label{f75}
  \end{minipage}
\end{figure}

  \begin{figure}
  \begin{minipage}[htbp]{0.48\linewidth}
    \centering
    \includegraphics[width=0.98\textwidth]{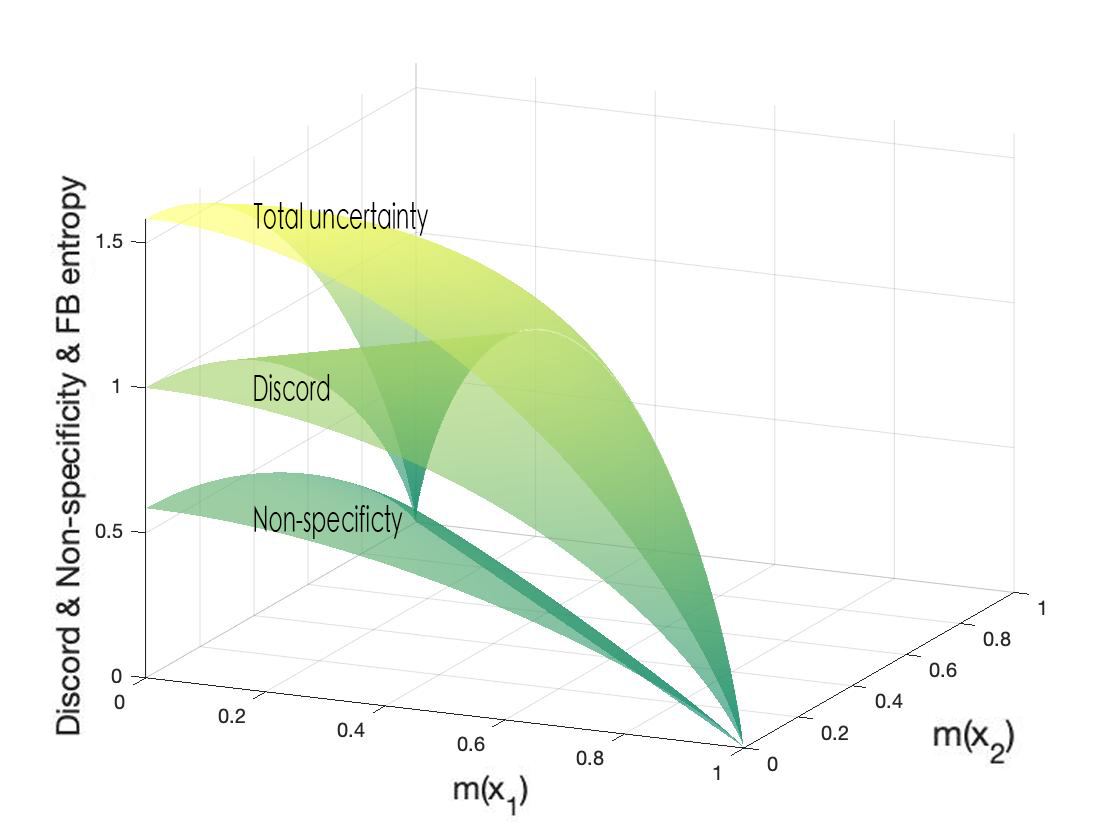}
    \caption{The relationship of TUM in Example \ref{e3}}
    \label{f72}
  \end{minipage}
  \begin{minipage}[htbp]{0.48\linewidth}
    \centering
    \includegraphics[width=0.98\textwidth]{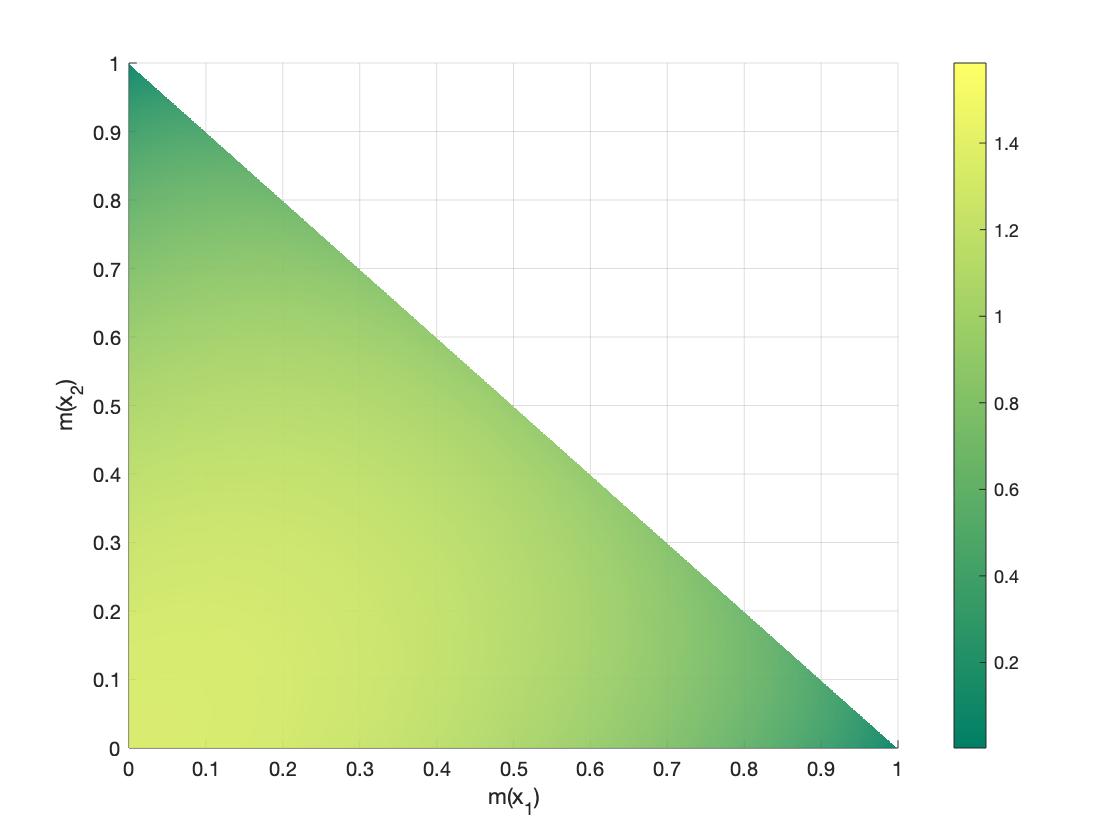}
    \caption{The top view of TUM in Example \ref{e3}}
    \label{f73}
  \end{minipage}
\end{figure}

\end{example}

\begin{proposition}[RB4($\heartsuit$)]\label{p10}
The proposed method is supposed to have corresponding model when meets more generalized theory than evidence theory.
\end{proposition}

\textbf{\emph{Proof.}}
FB entropy distributes mass functions uniformly to the power sets of their focal elements, because DSET uses power set $2^n$ to express information. As a generalization of DSET, the DSmT \cite{smarandache2006advances} proposed by Desert and Smarandache is to extend the power set $2^n$ to $U^n$. According to this idea, FB entropy can also measure the uncertainty of the assignment in DSmT, and only needs to uniformly distribute the mass functions to $U^n$ subsets. So FB entropy satisfies Proposition \ref{p10}.

So the FB entropy satisfies the Proposition \ref{p10}.

\qed

In this Section, $10$ requirements evaluate the general properties of FB entropy and prove its rationality. In particular, in terms of additivity, there was no previous total uncertainty measurement can complete the additivity verification on the basis of joint BPA. In the rest of the paper, we will further show the advantages of FB entropy.

\section{Advantages of FB entropy}\label{fbentropy2}

We make an intuitive comparison through several examples to show the advantages of FB entropy, which are not available in previous methods.

\subsection{View from combination rules: combination interval consistency}

Combination rule of Dempster (CRD) \cite{dempster2008upper} and disjointed combination rule (DCR) \cite{smets1993belief} are most widely used combination rules of normalized BPA. For BPAs $\mathbb{B}_1(2^\Theta)$ and $\mathbb{B}_2(2^\Theta)$ under discernment framework $\Theta$, the CRD $\mathbb{B}_{1\oplus2}$ and DCR  $\mathbb{B}_{1\circledtiny{$\cup$}2}$ are

\begin{equation}\label{crde}
m_{1\oplus 2}(F_i)=
\begin{cases}
K^{-1}\cdot \sum_{G_i\subseteq\Theta,H_i\subseteq\Theta}m_1(G_i)m_2(H_i)& \text{$G_i \cap H_i=F_i$}\\
0& \text{$F_i$ = $\varnothing$}
\end{cases},
\end{equation}

\begin{equation}
m_{1\circledtiny{$\cup$}2}(F_i)=\sum_{G_i\subseteq\Theta,H_i\subseteq\Theta,G_i \cup H_i=F_i}m_1(G_i)m_2(H_i),
\end{equation}

where $K=\sum_{H_i\cap G_i=\varnothing}m_1(G_i)m_2(H_i)$.

\begin{figure}
  \begin{minipage}[htbp]{0.48\linewidth}
    \centering
    \includegraphics[width=0.8\textwidth]{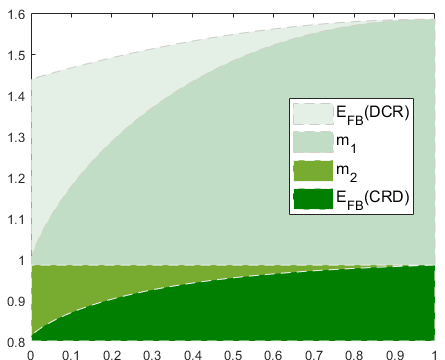}
    \caption{The change trend of FB entropy}
    \label{FB_combin}
  \end{minipage}
  \begin{minipage}[htbp]{0.48\linewidth}
    \centering
    \includegraphics[width=0.98\textwidth]{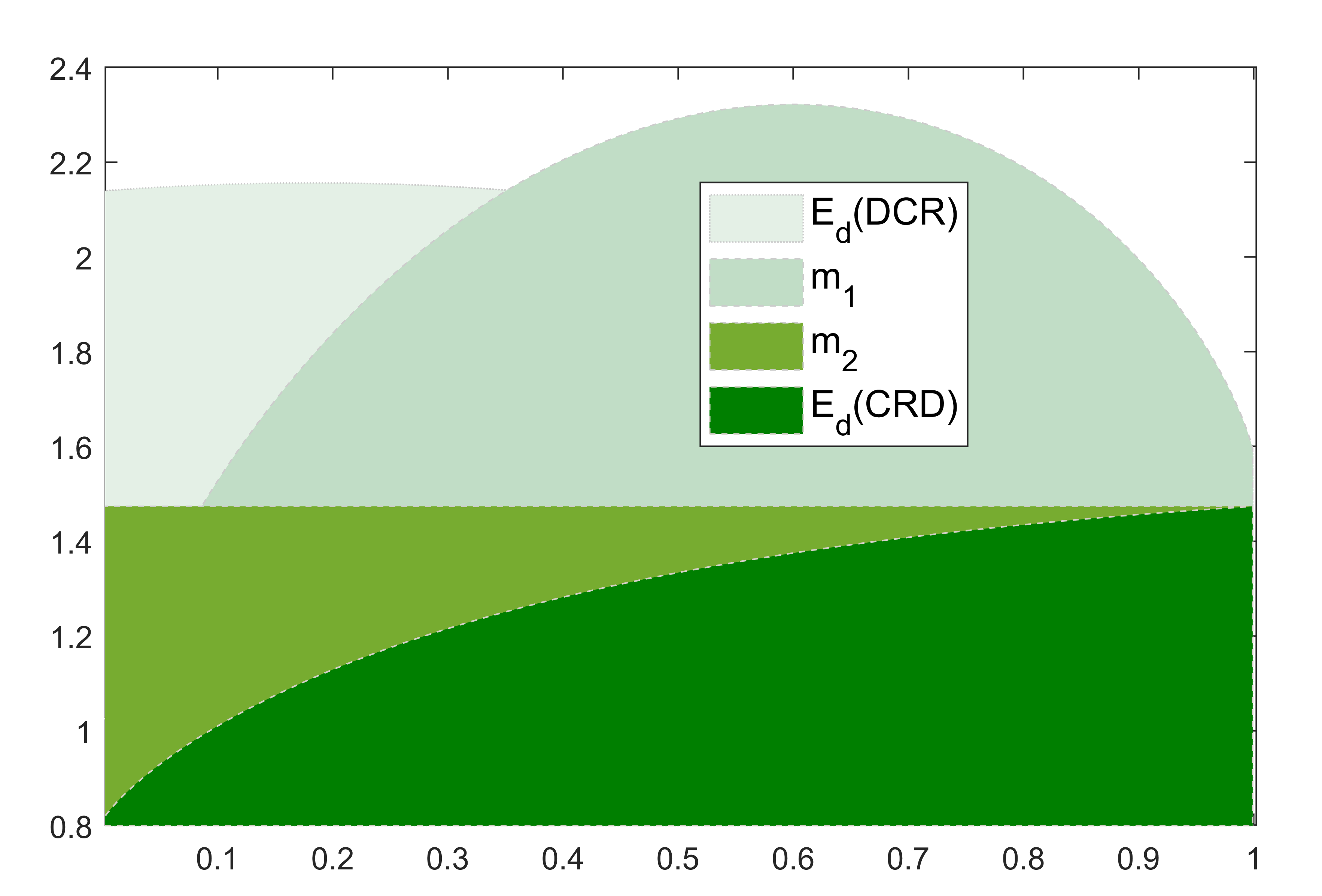}
    \caption{The change trend of Deng entropy}
    \label{Ed_combin}
  \end{minipage}
\end{figure}

CRD in DSET corresponds to the cross product in probability theory. The Shannon entropy of the probability after cross product is smaller than all original probabilities, which means that the uncertainty of the distribution are reduced after receiving new information. Intuitively, the uncertainty of the BPA after combination by CRD also should be reduced, so its total uncertainty measurement should satisfy $TUM(\mathbb{B}_{1\oplus2}) \leqslant \min\{TUM(\mathbb{B}_1),TUM(\mathbb{B}_2)\}$. DCR is a conservative combination rule. It assigns the mass functions of conflict evidence to their union, which is bound to cause more uncertainty. Therefore, the total uncertainty measurement of the BPA after using DCR should be larger and satisfies $TUM(\mathbb{B}_{1\circledtiny{$\cup$}2}) \geqslant \max\{TUM(\mathbb{B}_1),TUM(\mathbb{B}_2)\}$. So the total uncertainty measurement should satisfy the combination interval consistency, i.e., $\{TUM(\mathbb{B}_1),TUM(\mathbb{B}_2)\}\in[TUM(\mathbb{B}_{1\oplus2}),TUM(\mathbb{B}_{1\circledtiny{$\cup$}2})]$. For $2$ BPAs $\mathbb{B}_1=\{m(a)=m(b)=\frac{1-\frac{i}{1000}}{2},m(ab)=\frac{i}{1000}\}$ and $\mathbb{B}_2=\{m(a)=0.1,m(b)=0.7,m(ab)=0.2\}$, when $i$ from $0$ to $1$, Figure \ref{FB_combin} and \ref{Ed_combin} show the change trend of FB entropy and Deng entropy. From the Figures, it can be concluded that FB entropy meets the combination interval consistency but Deng entropy does not. So for this property, the measurement effect of FB entropy is better than Deng entropy.

\subsection{View from non-specificity: more rational measurement}

Non-specificity as a peculiar property of DSET, analyzing its uncertainty reasonably is significant. Besides the most well known Hartley entropy \cite{higashi1982measures}, Yang \textit{et al.} \cite{yang2016non} utilized belief interval to measure the non-specificity. In addition, common total uncertainty measurements can separate non-specificity, which are shown in Table\ref{tnon}. We evaluate these methods separately from qualitative and quantitative aspects.

\begin{table}[htbp!]
\caption{Non-specificity of common total uncertainty measurements}
\label{tnon}
\begin{center}
\small
\begin{tabular}{c|cccc}
  \Xhline{1.4pt}
Methods& \tabincell{c}{JS entropy~\&\\Pal \textit{et al.}'s entropy} & SU measurement & Deng entropy  &FB entropy \\
 \hline
Non-specificity&$\sum_{F_i\in 2^\Theta}m(F_i)\log |F_i|$&$\sum_{\theta_i\in\Theta}(Pl(\theta_i)-Bel(\theta_i))$&$\sum_{F_i\in 2^\Theta}m(F_i)\log (2^{|F_i|}-1)$&\tabincell{c}{$\sum_{\theta_i\in\Theta}BetP(\theta_{i})\log BetP(\theta_{i})$\\$-\sum_{F_i\in2^\Theta} m_{F}(F_i)\log m_{F}(F_i)$}\\
  \Xhline{1.4pt}
\end{tabular}
\end{center}
\end{table}

\textbf{Qualitative analysis:} SU measurement, JS entropy and other derivative methods \cite{Xue2021Interval} are to calculate the uncertainty of discord and non-specificity respectively and then add them up. In this way, discord and non-specificity are measured separately, which can reflect the relative uncertainty between different pieces of evidence. But for a BPA, we cannot know the proportion of discord and non-specificity in its total uncertainty. So logically speaking, these two parts should be divided from the total uncertainty measurement, instead of using these $2$ parts to form a method. From this point of view, Pal \textit{et al.}, Deng entropy and FB entropy are more reasonable.

\textbf{Quantitative analysis:} For a $4$-element discernment framework $X=\{a,b,c,d\}$ with $2$ evidence $\mathbb{B}_1(2^X)=\{m(ab)=m(bc)=m(cd)=m(ad)=\frac{1}{4}\}$ and $\mathbb{B}_2(2^X)=\{m(ab)=m(bc)=m(cd)=m(ad)=m(ac)=m(bd)=\frac{1}{6}\}$. According to non-specificity of them are shown in Table \ref{tnone}, the results of $\mathbb{B}_1(2^X)$ and $\mathbb{B}_2(2^X)$ measured by FB entropy are different. Although the belief intervals of their elements are all $[0,\frac{1}{2}]$, the probability ranges they can cover are not same. For example, $\mathbb{B}_2(2^X)$ can appear probability distribution $\{p(a)=\frac{1}{2},p(b)=\frac{1}{3},p(c)=\frac{1}{6}\}$, but $\mathbb{B}_1(2^X)$ only can reach $\{p(a)=\frac{1}{2},p(b)=\frac{1}{4},p(c)=\frac{1}{4}\}$. Only FB entropy can express this kind of difference, which proves its advantages in this aspect.

\begin{table}[htbp!]
\caption{Non-specificity of $\mathbb{B}_1(2^X)$ and $\mathbb{B}_2(2^X)$ }
\label{tnone}
\begin{center}
\small
\begin{tabular}{c|cccc}
  \Xhline{1.4pt}
Methods& \tabincell{c}{JS entropy~\&\\Pal \textit{et al.}'s entropy} & SU measurement & Deng entropy  &FB entropy \\
 \hline
$\mathbb{B}_1(2^X)$ &$1$&$2$&$1.5850$&$2.8554$\\
 \hline
$\mathbb{B}_2(2^X)$ &$1$&$2$&$1.5850$&$3.1133$\\
  \Xhline{1.4pt}
\end{tabular}
\end{center}
\end{table}

\subsection{View from physical model: Stronger ability to express information than Shannon entropy}

In Example \ref{e2}, we have used a physical model to show the difference between the maximum FB entropy and the maximum Shannon entropy. It is proved that the information volume of the maximum FB entropy under $n$-element discernment framework is equivalent to the information volume of the maximum Shannon entropy under $2^n-1$-events random variable. Expanding the cases in Example \ref{e2} can further prove the advantages of FB entropy. 

\begin{description}
\item[\textbf{Q}:] How many times inquiring can we find the champion at least?
\item[\textbf{Case1}:]We don’t know the exact number of champions.
\item[\textbf{Case2}:]We don't know the exact number of champions, but we know that the champions are in a certain half.
\item[\textbf{Case3}:]We don't know the exact number of champions, but we know that the champions are in a quarter of the population.
\item[\textbf{Case4}:]We don't know the exact number of champions, but we know that the champions are in a $\frac{1}{64}$ of the population.
\end{description}

\begin{figure}[htbp!]
\centering
\includegraphics[width=0.85\textwidth]{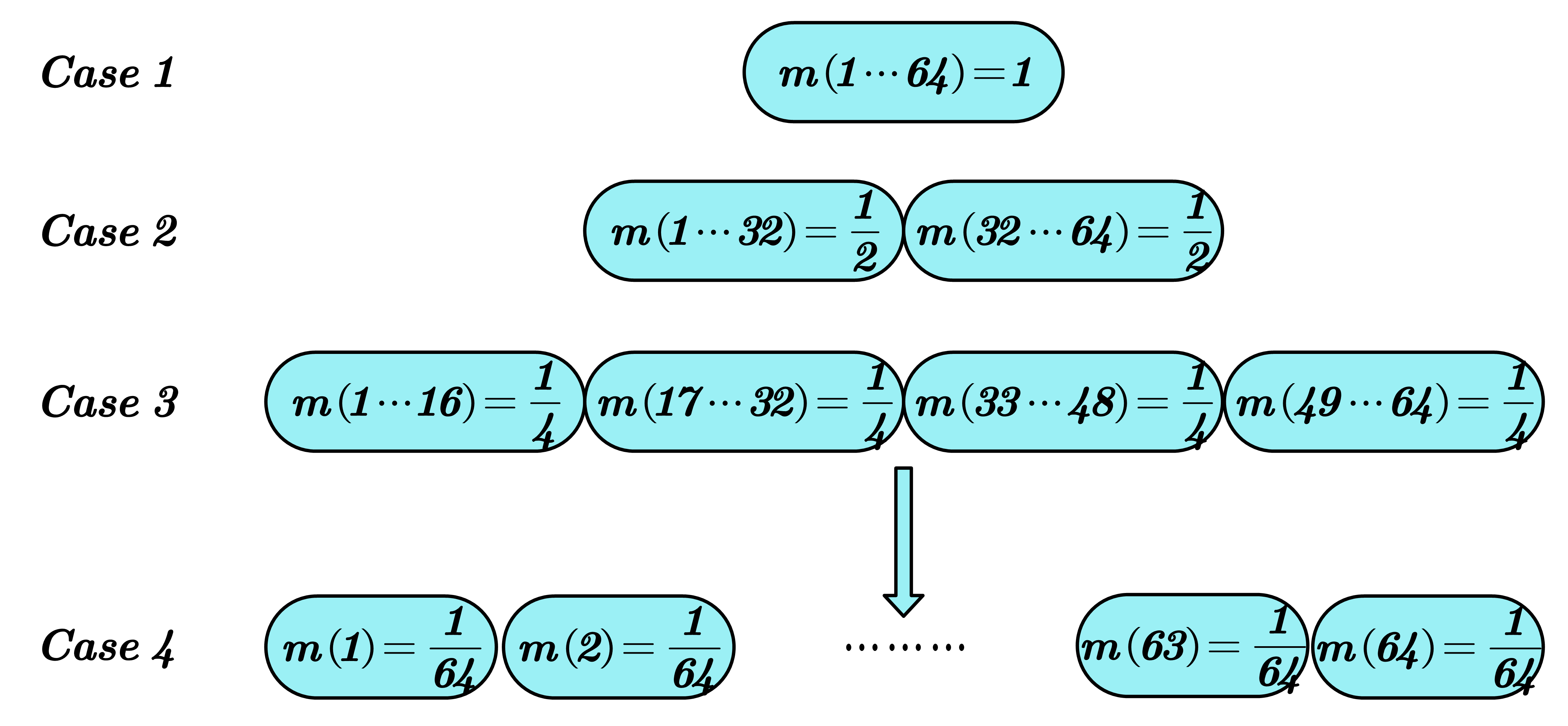}
\caption{The relationship of $4$ Cases.}
\label{ff}
\end{figure}

Among them, the BPA and probability distribution of Case $1$ and Case $4$ have shown in Example \ref{e2}, and Case $4$ also can be described as we knowing only have one champion. For Case $2$ and Case $3$, each of them can not be expressed by $1$ probability distribution, but BPAs $\mathbb{B}_{Case~2}=\{m(1\cdots 32)=m(33\cdots 64)=\frac{1}{2}\}$ and $\mathbb{B}_{Case~3}=\{m(1\cdots 16)=m(17\cdots 32)=m(33\cdots 48)=m(49\cdots 64)=\frac{1}{4}\}$ can express. The FB entropy of Case $2$ $E_{FB}(\mathbb{B}_{Case~2})\approx 33$, and in reality, we also need to inquire $33$ times to find all champions. First, we inquire $1$ time to find which half contain all champions, and then we can find  champions by inquiring all 32 people. For Case $3$, $E_{FB}(\mathbb{B}_{Case~2})\approx 18$, which also be consistent with physical model.  Based on the above, we can express the  the relationship of $4$ Cases in Figure \ref{ff}, which unifies the process of BPA degeneration to probability distribution and FB entropy degeneration to Shannon entropy. From the superiority of BPA compared to probability distribution, the superiority of FB entropy compared to Shannon entropy is inferred. At this point, FB entropy is better than all existing belief entropies.

\section{Conclusion}
This paper utilizes the fractal to simulate the process of pignistic probability transformation, which shows the process of lost information in transformation more intuitive. Based on the process, we propose the fractal-baed belief entropy to measure the BPA's total uncertainty. After 10 required evaluations, we prove that FB entropy can reasonably measure the uncertainty of BPA. In addition, we prove its superiority from $3$ aspects: combination rule, non-specificity and physical model. Based on above, the contributions of paper are summarized as follows:

\begin{itemize}
\item[$\bullet$] [\textbf{Process of probability transformation:}] We consider probability transformation as a process, and propose a possible transformation process of PPT and PMT. This idea provides a new perspective to evaluate probability transformation, so that result orientation is no longer the only evaluation criterion. 
\item [$\bullet$][\textbf{Total uncertainty measurement of BPA:}] Based on fractal, we propose the FBBPA and substitute it into Shannon entropy to define the FB entropy. After evaluation, FB entropy can not only measure total uncertainty of BPA reasonably, but satisfy the additivity, which realize the corresponding with Shannon entropy.
\item[$\bullet$] [\textbf{Combination rule interval:}] Based on the CRD and the DCR, we propose the combination interval consistency and prove the FB entropy is better than Deng entropy in this property.
\item[$\bullet$] [\textbf{Discord and non-specificity measurement:}] Since PPT is the end point of the fractal method, we substitute it into Shannon entropy as discord measurement. Through qualitative and quantitative analysis, we prove that FB entropy is superior to all previous uncertainty measurement methods in this aspect.
\item[$\bullet$] [\textbf{Physical model consistency: }]  As a generalization of Shannon entropy, FB entropy can not only degenerate into Shannon entropy when the input is probability distribution, but correspond to Shannon entropy in the physical model of maximum entropy as well.
\end{itemize}

In future research, this work can be further extended from three directions. (1) In terms of probability transformation, more probability transformation methods can be simulated based on the proposed process model. (2) In terms of uncertainty measurement, this fractal-based measurement method can be applied to more uncertainty theories. (3) In DSET, FB entropy can be applied to solve practical problems such as information fusion, decision making and fault diagnosis.

\section*{Declaration of interests}
The authors declare that they have no known competing financial interests or personal relationships that could have appeared to influence the work reported in this paper.

\section*{Acknowledgment}

The work is partially supported by National Natural Science Foundation of China (Grant No. 61973332), JSPS Invitational Fellowships for Research in Japan (Short-term). Thanks to the reviewers' valuable comments, which significantly improved the quality of the paper. Thanks to colleagues in the Information Fusion and Intelligent System Laboratory for their help and support.

\bibliography{mybibfile}

\end{document}